\documentclass[twocolumn]{aastex63}

\usepackage{amsmath}
\usepackage{bm}
\usepackage{graphicx}
\usepackage{float}
\usepackage{multirow}
\usepackage{makecell}
\usepackage{booktabs}
\usepackage{hyperref}
\usepackage{ulem}

\newcommand{\Mjup}{\ensuremath{M_\mathrm{Jup}}}
\newcommand{\Bpic}{\ensuremath{\beta \ \text{Pic}}}

\newcommand{\mps}{\ensuremath{\text{m s}^{-1}}}
\newcommand{\dra}{\ensuremath{\Delta \mathrm{RA}}}
\newcommand{\ddec}{\ensuremath{\Delta \mathrm{DEC}}}
\newcommand{\Mtot}{\ensuremath{M_\mathrm{Tot}}}

\received{June 30, 2020}
\accepted{September 17, 2020}
\submitjournal{AJ}

\shorttitle{Dynamical Masses of the $\beta$ Pic System Through GP}
\shortauthors{Vandal, Rameau and Doyon}

\begin{document}

\title{Dynamical Mass Estimates of the $\beta$ Pictoris Planetary System\\
Through Gaussian Process Stellar Activity Modelling}

\correspondingauthor{Thomas Vandal}
\email{vandal@astro.umontreal.ca}

\author[0000-0002-5922-8267]{Thomas Vandal}
\affiliation{Institut de Recherche sur les Exoplan\`etes,
D\'epartement de physique, Universit\'e de Montr\'eal,
CP 6128 Succ. Centre-ville, H3C 3J7, Montr\'eal, QC, Canada}
\affiliation{Department of Physics, McGill University,
3600 rue University, H3A 2T8, Montr\'eal, QC, Canada}

\author[0000-0003-0029-0258]{Julien Rameau}
\affiliation{Universit\'e Grenoble Alpes/CNRS,
Institut de Plan\'etologie et d'Astrophysique de Grenoble,
38000 Grenoble, France}
\affiliation{Institut de Recherche sur les Exoplan\`etes,
D\'epartement de physique, Universit\'e de Montr\'eal,
CP 6128 Succ. Centre-ville, H3C 3J7, Montr\'eal, QC, Canada}

\author[0000-0001-5485-4675]{Ren\'e Doyon}
\affiliation{Institut de Recherche sur les Exoplan\`etes,
D\'epartement de physique, Universit\'e de Montr\'eal,
CP 6128 Succ. Centre-ville, H3C 3J7, Montr\'eal, QC, Canada}

\begin{abstract}
Nearly 15 years of radial velocity (RV) monitoring and direct imaging enabled the detection 
of two giant planets orbiting the young, nearby star $\beta$ Pictoris. The $\delta$ Scuti pulsations of the star, overwhelming planetary signals, need to be carefully suppressed. In this work, we independently revisit the analysis of the RV data following a different approach than in the literature to model the activity of the star. We show that a Gaussian Process (GP) with a stochastically driven damped harmonic oscillator kernel can model the $\delta$ Scuti pulsations. It provides similar results as parametric models but with a simpler framework, using only 3 hyperparameters. It also enables to model poorly sampled RV data, that were excluded from previous analysis, hence extending the RV baseline by nearly five years. Altogether, the orbit and the mass of both planets can be constrained from RV only, which was not possible with the parametric modelling. To characterize the system more accurately, we also perform a joint fit of 
all available relative astrometry and RV data. Our orbital solutions for \Bpic{} b 
favour a low eccentricity of $0.029^{+0.061}_{-0.024}$ and a relatively short period of 
$21.1^{+2.0}_{-0.8}$~yr. The orbit of \Bpic{} c is eccentric with $0.206^{+0.074}_{-0.063}$ with a period of $3.36\pm0.03$~yr. We find model-independent masses of $11.7\pm1.4$ and 
$8.5\pm0.5$~\Mjup{} for \Bpic{} b and c, respectively, assuming coplanarity. The mass of \Bpic{} b is consistent with the hottest start evolutionary models, at an age of $25\pm3$ Myr. A direct direction of \Bpic{} c would provide a second calibration measurement in a coeval system.
\end{abstract}

%% Replaced by UAT Concepts during submission process
%% Uncomment only for Arxiv

\section{Introduction}\label{sec:intro}

Exoplanets directly detected must rely on theoretical models to determine their masses from their luminosity and age. Not only is age difficult to evaluate \citep{soderblom_age_2013}, but evolutionary models are also uncalibrated at young ages ($\le1$~Gyr) and low masses ($\le70~$\Mjup). Model-independent mass measurements must be obtained to test these models, which ultimately will provide confidence in inferred masses for all directly imaged planets. Absolute astrometry with exquisite precision is amenable to achieve this goal. Doppler spectroscopy of planet host stars can also determine the planet masses, biased with the orbit inclination which can be constrained from relative astrometry to overcome this bias. However, directly imaged planets are usually found at long periods around young stars since they are bright and spatially resolved with current instruments \citep[see][and references therein]{bowler_planets_2014}. Radial velocity (hereafter RV) data must therefore be obtained over a long timescale. These data are however severely plagued by strong stellar activity signal due to spots, plages, and pulsations \citep{galland_extrasolar_2005,lagrange_feasibility_2013,hillenbrand_limits_2014}. Planet detection is thus very challenging on both observational and data analysis sides. However, these limitations are less severe for low-mass star and high-mass brown-dwarf companions to stars, enabling independent mass and luminosity measurements using a combination of relative, absolute astrometry, and/or radial velocity data. Evolutionary model calibration is therefore being undertaken for these objects. \citep{crepp_trends_2012,ryu_high-contrast_2016,dupuy_individual_2017}. In the planetary mass regime, reduced signal amplitudes (RV, astrometry, contrast) make this task much harder.

$\beta$ Pictoris (hereafter \Bpic{}) is a nearby ($19.44 \pm 0.05$ pc, \cite{van_leeuwen_validation_2007}), A6V \citep{gray_contributions_2006}, intermediate-mass ($1.80^{+0.03}_{-0.04}$ M$_\odot$ 
\cite{wang_orbit_2016}) star. \Bpic{} has been classified as a $\delta$ Scuti variable \citep{koen__2003} characterized by pulsation frequencies in the range 22.81-75.68 d$^{-1}$ \citep{mekarnia__2017,zwintz_revisiting_2019,zieba_transiting_2019}. \Bpic{} is a member of the $\beta$ Pictoris moving group \citep{zuckerman__2001}, with recent age estimates of $24\pm3~$Myr \citep{bell_pre-main-sequence_2014} and $26\pm3~$Myr \citep{nielsen_dynamical_2016} using isochrones fitting, or $25\pm3~$Myr by measuring the lithium depletion boundary on group members \citep{messina_rotation-lithium_2016}. \Bpic{} is surrounded by a complex environment. The large circumstellar disk of gas and debris \citep{smith_circumstellar_1984,dent_gaspsherschel_2013} has a primary component seen almost edge-on and an inner warp component titled by $+4.0\pm0.6^\circ$ and further inclined by $6\pm1^\circ$ \citep{burrows_hst_1995,mouillet_planet_1997, golimowski_hubble_2006, ahmic_dust_2009, lagrange_position_2012}. Falling evaporating bodies or exocomets were evidenced \citep{lagrange__1996,kiefer_two_2014, zieba_transiting_2019}. Both warp and comets were modelled with gravitational perturbations by planet(s) within the system \citep{beust_beta_1990, beust_mean-motion_1996,beust_beta_1998, mouillet_planet_1997}. Two giant planets were discovered within the system: \Bpic{} b by direct imaging \citep{lagrange_giant_2010} and \Bpic{} c with RV \citep{lagrange_evidence_2019}. Monitoring the stellar RVs since 2008 (2003) up to 2018 combined with a multiparametric sinusoidal fit for the $\delta$ Scuti pulsations led to a residual noise of $10~\mathrm{m}~\mathrm{s}^{-1}$, allowing for the detection of \Bpic{} c. The system is therefore a rare case for which independent mass measurement can be foreseen, as well as a deep understanding of a complex planetary system.

%\explain{Edited this paragraph to mention results with and without absolute astrometry for Nielsen et al. - T.V.}
Simultaneous imaging of \Bpic{} b and the disk proved that \Bpic{} b is not coplanar with the main disk but consistent for being responsible for the inclined warp  \citep{lagrange_position_2012}. Monitoring the motion of \Bpic{} b with direct imaging has led to various constraints on its orbit. The inclination is now well constrained, excluding perfectly edge-on orbit, at $i=89.04\pm0.04^\circ$ \citep{wang_orbit_2016,lagrange_post-conjunction_2019,nielsen_gemini_2020, nowak_peering_2020}.
Period and eccentricity were also constrained with direct imaging.
\cite{lagrange_post-conjunction_2019} reported $P_\mathrm{b} = 20.29^{+0.86}_{-1.35}$ yr and $e_\mathrm{b} = 0.01^{+0.03}_{-0.01}$ when using relative astrometric data only from VLT instruments (NaCo, SPHERE) while
\cite{nielsen_gemini_2020} obtained $P_\mathrm{b} = 21.3^{+2.2}_{-1.0}$ yr and $e_\mathrm{b} = 0.038^{+0.063}_{-0.029}$ with all relative astrometric measurements except those from VLT/SPHERE.
However, the inclusion of absolute astrometry from \textit{Hipparcos} and \textit{Gaia} in the analysis by \cite{nielsen_gemini_2020} lead to significantly higher period and eccentricity values, namely $P_\mathrm{b} = 24.4^{+1.0}_{-1.5}$ yr and $e_\mathrm{b} = 0.12^{+0.04}_{-0.03}$.
An analysis of the absolute astrometric motion of the star allowed \cite{snellen_mass_2018} to directly measure a planetary mass of $M_\mathrm{b} = 11 \pm 2$ \Mjup{}; \citet{dupuy_model-independent_2019} found a mass of  $M_\mathrm{b} = 13 \pm 3$ \Mjup{} with additional corrections to the same data and further including relative astrometry and radial velocity data and fitting as well for the mass of the star.
Analysis of the stellar RVs using VLT relative astrometric priors yielded $M_\mathrm{b} \sim 10$ \Mjup{} \citep{lagrange_evidence_2019}. Lastly, a joint analysis of direct imaging, RV and astrometric data accounting for both planets resulted in a mass of $8.0 \pm 2.6$ 
\Mjup{} for \Bpic{} b \citep{nielsen_gemini_2020}.
This is lower than previous estimate, but still consistent within $1\sigma$. These last two mass measurements account for the presence of a second giant planet in the system.
The addition of a VLTI/GRAVITY relative astrometric point with exquisite precision of $80~\mu\mathrm{as}$, but without the RVs and without accounting for a second planet, led \cite{nowak_peering_2020} to measure $M_\mathrm{b}=12.7\pm2.2$ \Mjup{} while excluding circular orbit ($e_\mathrm{b}=0.15^{+0.05}_{-0.04}$).

From stellar RVs only, the mass and period of \Bpic{} c were estimated to $M_\mathrm{c} \sim 8.9$ \Mjup{} (assuming the same inclination as \Bpic{} b) and $P_\mathrm{c} \sim 1220$ d 
\citep{lagrange_evidence_2019}, while \citet{nielsen_gemini_2020} found $M_\mathrm{c}=9.18^{+0.96}_{-0.87}~$\Mjup{}, $P_\mathrm{c} =1238^{+7}_{-11}$ d, and $e_\mathrm{c}=0.24^{+0.11}_{-0.10}$.

RV data are therefore of prime importance when fitting for the masses of both planets, the absolute astrometry being still plagued by imperfect knowledge, bias, and required corrections of Gaia DR2 data onto bright stars \citep{brown_gaia_2018}. For \Bpic{}, the $\delta$ Scuti pulsations result in RV noise with a semi-amplitude 
of $\sim 500$ m s$^{-1}$, while \Bpic{} b and c have expected semi-amplitudes of 
$K_\mathrm{b} \sim 70$ m s$^{-1}$ and $K_\mathrm{c} \sim 120$ m s$^{-1}$, respectively. The main 
challenge is therefore to properly model and subtract the stellar activity, thus 
revealing planetary signals in the RV data.
    
A useful and increasingly used method to mitigate the stellar noise in the RV data of an 
active star is to use Gaussian processes (GP). There are several 
successful attempts to use this method in the literature (e.g., 
\cite{haywood_planets_2014, grunblatt_determining_2015, cloutier_characterization_2017, 
cloutier_radial_2017}). However, these examples are generally limited to less massive
stars from the field for which the stellar activity has fewer components and less temporal variations 
than for young stars like \Bpic{}.
    
Here, we investigate possible improvements of the modelling of stellar 
activity in the RV data of \Bpic{} in a Bayesian framework. We explore 
GPs for correlated Gaussian noise modelling over traditional parametric approaches from the literature 
\citep{lagrange_evidence_2019}. This investigation leads to a measurement of the orbits and masses of both \Bpic\,b and c from RV data only, an improvement over the literature that was unable to constrain the orbital parameters and mass of \Bpic{} b from solely the same data. We then present a new joint analysis of all RV and astrometric data available, including RV points 
between 2003 and 2008 that were previously ignored in the literature. This analysis leads to new mass estimates for both planets.

The paper is structured as follows. Section~\ref{sec:data}     
gives an overview of all archival data used for the analysis. In \S~\ref{sec:act}, we outline 
the main elements of the GP method used to define a new stellar activity model for 
\Bpic{}, followed in \S~\ref{sec:orbit} by a detailed orbit fitting analysis using RV alone, then adding astrometry data. A discussion of these results in presented  \S~\ref{sec:discussion} followed by a summary and conclusions in \S~\ref{sec:conclusion}.

\section{Literature data}\label{sec:data}

In this work, we made use of published data that monitored the system over nearly the last 15 years. We were agnostic to the type and origin of the data but deliberately excluded absolute astrometric measurements from \textit{Gaia}. Their analysis requires multiple known correction factors to compensate the star being very bright for the instrument \citep{brown_gaia_2018} but also still unknown analysis steps to provide accurate measurements. We defer their inclusion into this type of work once the \textit{Gaia} data release will be secured for bright stars.

\subsection{ASTEP Photometry Timeseries}\label{sec:data_astep}

\begin{figure}[t!]
    \centering
    \includegraphics[width=\linewidth]{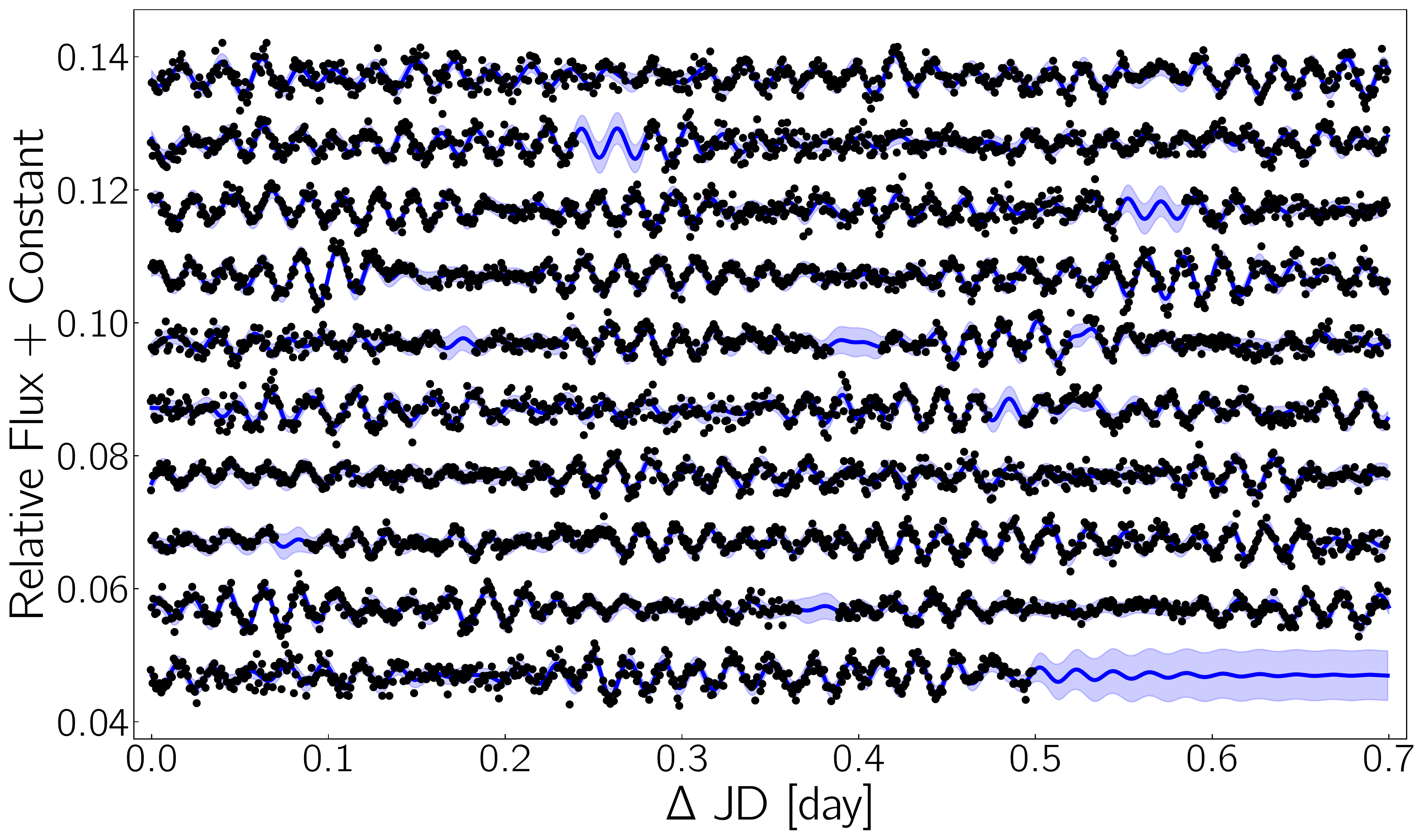}
    \caption{ASTEP photometric light curve of \Bpic{}. The black dots are the observations between 7 (top left) and 13 (bottom right) June 2017. The data is split in segments of length $\Delta$JD = 0.7 day and a constant offset is added to each segment to show the pulsations clearly. The solid blue line is the fit from an SHO GP and the shaded area is the 1$\sigma$ confidence interval.}
    \label{fig:astep_photo}
\end{figure}

We had access to 6 358 photometric observations of \Bpic{} from the Antarctica 
Search for Transiting Extrasolar Planets (ASTEP) 400 mm telescope 
\citep{crouzet_four_2018}, obtained between 7 and 13  June 2017 (JD~=~2~457~910 to 
JD~=~2~457~917), with a sampling rate of $\sim 1000 \ \text{d}^{-1}$ (\citealt{mekarnia__2017}, private communication). This light curve 
is shown in Figure~\ref{fig:astep_photo}, along with a GP fit 
explained in \S~\ref{sec:act_photo}. These observations were part of a longer 
monitoring campaign of \Bpic{} (March to September 2017, \citealt{mekarnia__2017}). For 
this longer light curve, a periodogram analysis allowed \cite{mekarnia__2017} to detect 31 
$\delta$ Scuti pulsation modes and to model the photometric stellar activity as a sum 
of sine waves. Since all of the 31 detected frequencies were in the 34.76-75.68 d$^{-1}$ 
interval, the fact that we only had access to a small portion of the photometric time 
series (6.8 d) should not prevent us from characterizing the activity of the star.
    
\begin{figure}
    \centering
    \includegraphics[width=\linewidth]{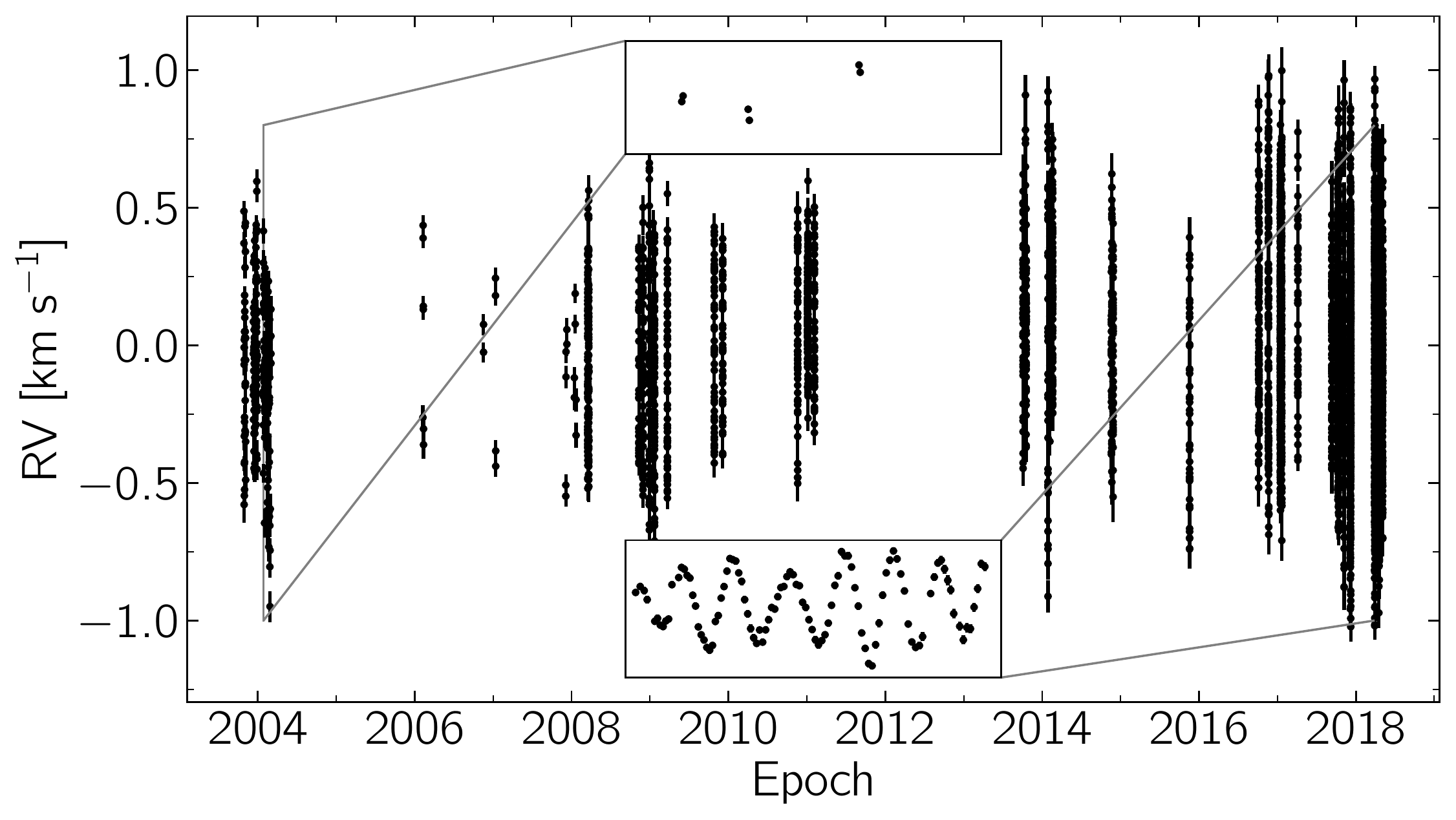}
    \caption{HARPS RV timeseries of \Bpic{}. The inset on the bottom shows an example of the pulsations causing the high-amplitude noise in the RV data. The top inset shows how the sparse pre-2008 sampling hides these pulsations.}
    \label{fig:harps_rv}
\end{figure}

\subsection{HARPS RV Timeseries}\label{sec:data_harps}

\cite{lagrange_constraints_2012,lagrange_evidence_2019} presented 5362 radial velocity measurements 
of \Bpic{} obtained with the High Accuracy Radial velocity Planet Searcher (HARPS, 
\cite{mayor_setting_2003}). These RV measurements, shown in Figure~\ref{fig:harps_rv} and not corrected for stellar activity, were 
computed with the SAFIR package \citep{galland_extrasolar_2005}.
These measurements, taken between October 2003 and May 2018, were sampled 
with two different methods. The sampling is much denser after March 2008, and reveals 
high-frequency variations in the RV data, with a similar timescale to the $\delta$ Scuti pulsations detected in the photometry (\S~\ref{sec:data_astep}), that were not visible before that 
date \citep{lagrange_constraints_2012}. The two insets in 
Figure~\ref{fig:harps_rv} show example of these two different sampling methods (pre-2008 and 
post-2008).
The poor sampling from before 2008 needs to be considered carefully when analyzing this 
data, especially if we attempt to model the high-frequency stellar activity in the radial velocities.
One option is to neglect pre-2008 data and use only the well-sampled post-2008 data 
\citep{lagrange_evidence_2019, nielsen_gemini_2020}. However, the observation baseline 
of the RV data is shorter than the period of \Bpic{} b, so 5 additional years of data, leading to 
more than 14 years of RV coverage, would provide valuable information about the system.
This is addressed in \S~\ref{sec:act_rv}, where we discuss the 
inclusion of pre-2008 data.

\subsection{Relative Astrometry}

We use all relative astrometry data available for \Bpic{}. This 
includes 
nine epochs from VLT/NaCo \citep{lagrange_giant_2010,currie_5_2011,chauvin_orbital_2012}, 
seven epochs from Gemini-South/NICI \citep{nielsen_gemini_2014}, 
two epochs from Magellan/MagAO \citep{morzinski_magellan_2015}, 
fifteen epochs from Gemini-South/GPI \citep{wang_orbit_2016,nielsen_gemini_2020} and 
twelve epochs from VLT/SPHERE \citep{lagrange_post-conjunction_2019}. These measurements are 
all shown in Figure~\ref{fig:joint_fit}.

We also include the single epoch of relative astrometry obtained with VLTI/GRAVITY by 
\cite{nowak_peering_2020} on September 22 2018. 
This data point, obtained by averaging 17 exposure files, is given by 
the bivariate normal distribution $(\dra{}, \ddec{}) \sim \mathcal{N}(\bm{\mu}, \bm{\Sigma})$. 
$\bm{\mu}$ is the mean relative planet-to-star position and $\bm{\Sigma}~=~\mathrm{Cov}(\dra{},\ddec{})$, the covariance matrix of all exposure files, gives the $1\sigma$ 
confidence interval. The mean coordinates and their covariance, as reported by \cite{nowak_peering_2020}, are
\begin{equation}\label{eq:gravity}
    \left\{\begin{array}{ll}
        \bm{\mu} = \begin{bmatrix} \dra{}\\ \ddec{} \end{bmatrix}
            = \begin{bmatrix} 68.48\\ 128.31 \end{bmatrix} \mathrm{mas}
            \vspace{3mm}\\
        \bm{\Sigma} = \begin{bmatrix} 
                        0.0027 & -0.0035\\
                        -0.0035 & 0.0045
                      \end{bmatrix} \mathrm{mas}^2.
    \end{array}\right.
\end{equation}
This yields a precision more than an order of magnitude better than the other relative 
astrometry measurements mentioned above. The GRAVITY data point is shown in the inset of 
Figure~\ref{fig:joint_fit}.

\section{Stellar Activity Modelling}\label{sec:act}

\subsection{ASTEP Photometry}\label{sec:act_photo}
\cite{mekarnia__2017} detected 31 $\delta$ Scuti pulsation frequencies between 34.76 and 75.68 
d$^{-1}$ in the ASTEP photometric light curve of \Bpic{}, with a peak at 47.43 d$^{-1}$ (period 
of 30.4 min.). Similar high-frequency pulsations appear in the RV data 
\citep{lagrange_constraints_2012, lagrange_evidence_2019}, and their amplitude is far greater 
than the expected signals from \Bpic{} b and c. Efficiently modelling and removing this stellar
activity in the RV data is crucial if we intend to detect Keplerian planetary signals. Since the 
RV and photometric datasets exhibit similar pulsation patterns, we first train a null-mean GP 
on the ASTEP photometry. We use the \texttt{celerite} Python package 
\citep{foreman-mackey_fast_2017} to compute the GP models used in the following analysis. 
To obtain posterior 
distributions of the GP hyperparameters, we use the affine-invariant Markov chain Monte-Carlo 
(MCMC) sampler from \texttt{emcee} \citep{foreman-mackey_emcee:_2013}. We use 128 walkers and ensure that all chains are longer than 40 times the integrated autocorrelation time. 
Once we have a distribution of the GP hyperparameters from photometry, we can 
apply it as a prior when fitting stellar activity in the RV data.

We first explore the use of a Quasi-Periodic (QP) kernel typically employed to model stellar 
rotation for stars quieter than \Bpic{} \citep{haywood_planets_2014, rajpaul_gaussian_2015, grunblatt_determining_2015, cloutier_characterization_2017, angus_inferring_2018}. We refer to \cite{foreman-mackey_fast_2017} for an explicit definition of the QP kernel, both in the \texttt{celerite} framework (Equation 56) and in its more classical form (Equation 55).
We find that this kernel is able to properly model the pulsations of \Bpic{}, recovering the main 
period with $P = 30.4$ min. 

\begin{figure}
    \centering
    \includegraphics[width=0.95\linewidth]{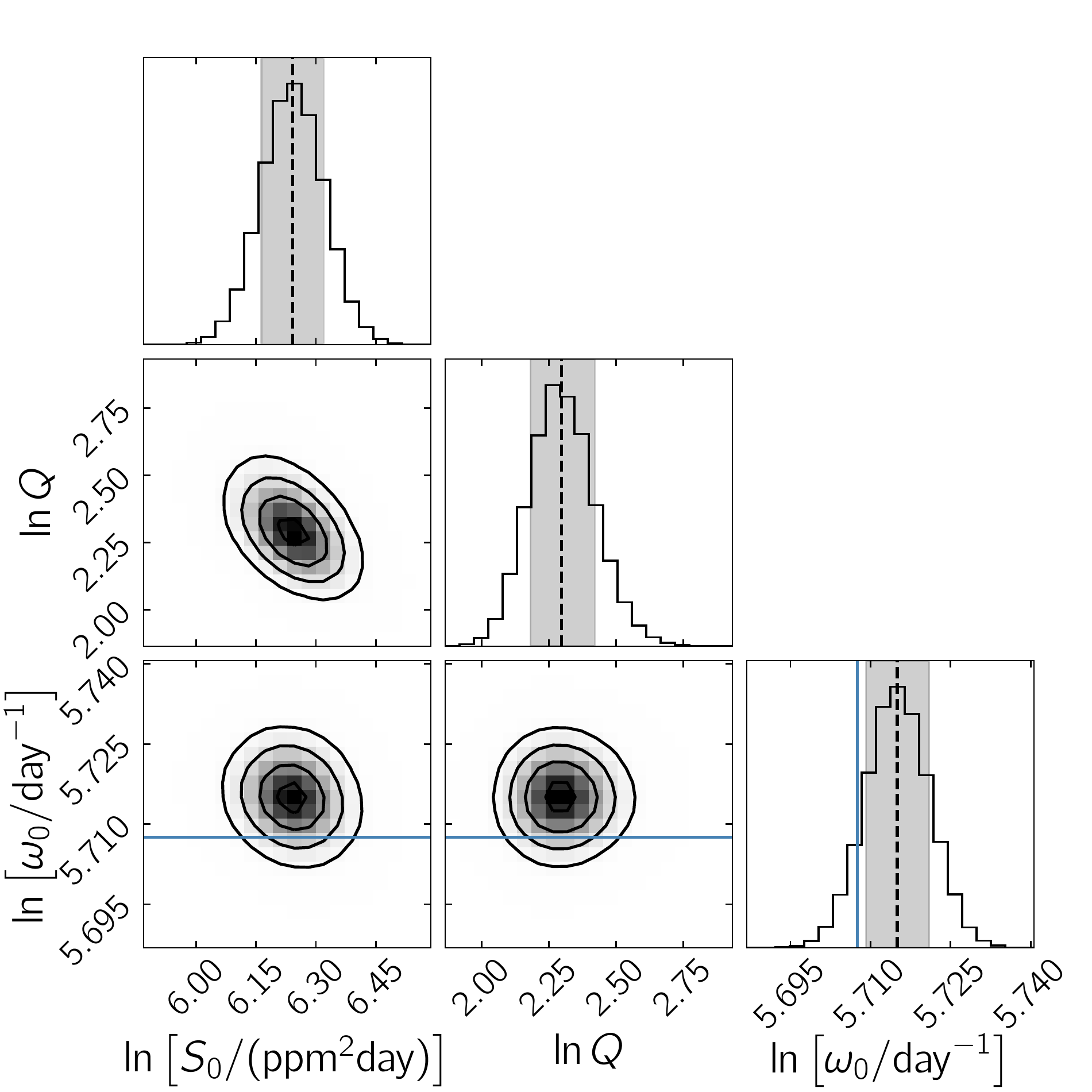}
    \caption{Posterior distribution of the SHO GP hyperparameters from the fit on the ASTEP photometry. Broad, uniform priors were used on all three hyperparameters. The dashed lines and grey shaded areas show the median and the $1\sigma$ interval for each hyperparameter. The solid blue line shows the main frequency detected by \cite{mekarnia__2017}.}
    \label{fig:photo_sho_post}
\end{figure}

\cite{foreman-mackey_fast_2017} introduced a covariance function that represents a 
stochastically driven damped simple harmonic oscillator (SHO). This kernel is defined as:
\begin{multline}\label{eq:SHO}
    k_\mathrm{SHO}(\tau; S_0, Q, \omega_0) = 
    S_0 \omega_0 Q \exp{\left(-\frac{\omega_0 \tau}{2 Q}\right)}\\
    \times \begin{cases}
        \cosh{(\eta \omega_0 \tau)} + \frac{1}{2 \eta Q}\sinh{(\eta \omega_0 \tau)}, & 0<Q<1/2\\
        2 (1 + \omega_0 \tau),                                                       & Q=1/2\\
        \cos{(\eta \omega_0 \tau)} + \frac{1}{2 \eta Q}\sin{(\eta \omega_0 \tau)}, & Q>1/2\\
    \end{cases},
\end{multline}
where $\eta = \left|1-(4Q^2)^{-1}\right|^{1/2}$, $\omega_0$ is the frequency of the undamped 
oscillator, $Q$ is the quality factor and $S_0$ is proportional to the power at $\omega = \omega_0$, $S(\omega_0)~=~\sqrt{2/\pi} S_0 Q^2$. $\tau = \left|t_n-t_m\right|$ is the absolute time difference between any two observations taken at times $t_n$ and $t_m$. When varying the quality factor, the SHO GP can provide an effective model for a large variety of phenomena. In the large $Q$ limit, it can mimic 
quasi-periodic pulsations. Moreover, the spectrum of an SHO covers a certain range of frequencies that depends on its hyperparameters. This ability to account for a wide range of frequencies provides some physical justification to adopt such a kernel in the context of a pulsating star like \Bpic{}. 

We find that the SHO GP models the $\delta$ Scuti pulsations of \Bpic{} adequately, yielding an $\omega_0$ value corresponding to $P \sim 30$ min. The posterior distribution of hyperparameters is shown in Figure~\ref{fig:photo_sho_post}. We note that the BIC favours the SHO kernel to the QP kernel very strongly, with $\Delta\mathrm{BIC} = 227$.

We consider both kernels in the following analysis of the RV data since both sampling and baseline are different from that of the light curve, possibly affecting their responses to the fit.

\subsection{HARPS RV}\label{sec:act_rv}

\begin{figure}
    \centering
    \includegraphics[width=0.95\linewidth]{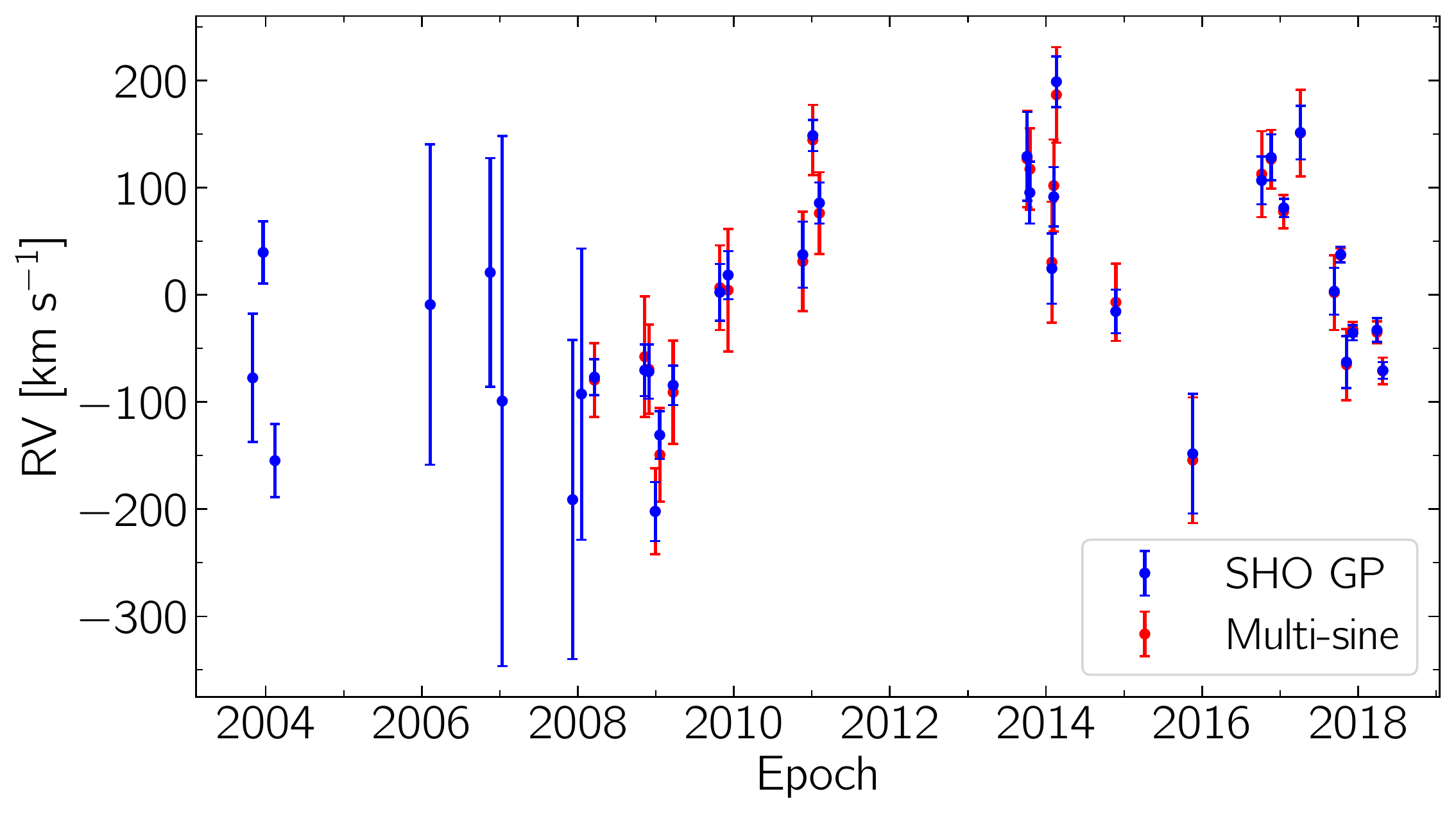}
    \caption{RV offsets for each group. The multi-sine offsets from \cite{lagrange_evidence_2019} are shown in red and the SHO GP offsets from this work are shown in blue. All values for post-2008 offsets are consistent within $1\sigma$ and the uncertainties on the GP  offsets tend to be smaller than on the multi-sine ones by a factor of 1.5.}
    \label{fig:gp_multisin_compare}
\end{figure}

Typically, when a GP is used to model stellar activity in RV data, the GP and the Keplerian orbits 
are jointly fit, i.e. the mean function of the GP is the Keplerian model and they are modelled 
all at once on the entire dataset. This method has been successfully used in the past (e.g. 
\cite{haywood_planets_2014, rajpaul_gaussian_2015, grunblatt_determining_2015, cloutier_characterization_2017}).
However, the stars studied were typically much quieter than \Bpic{}. 
For this reason, we investigate an alternative approach similar to the one used by 
\cite{lagrange_evidence_2019}: we model the activity locally by splitting the data into subsets 
corresponding to continuous observing sequences (hereafter referred to as groups). 
Within these groups the maximum time separation between observations is 8 days. 
For each group, we use a GP model trained on the photometry, i.e. using posterior distributions from photometry as priors, with a constant mean. 
By letting all parameters vary in our MCMC, i.e. the GP hyperparameters and the constant mean, we can obtain an offset with an uncertainty (from the MCMC posterior distribution) for each group. 
These offsets thus represent the RV of the star at each epoch, and we can then fit them with a Keplerian model.

\begin{figure}
    \centering
    \includegraphics[width=0.95\linewidth]{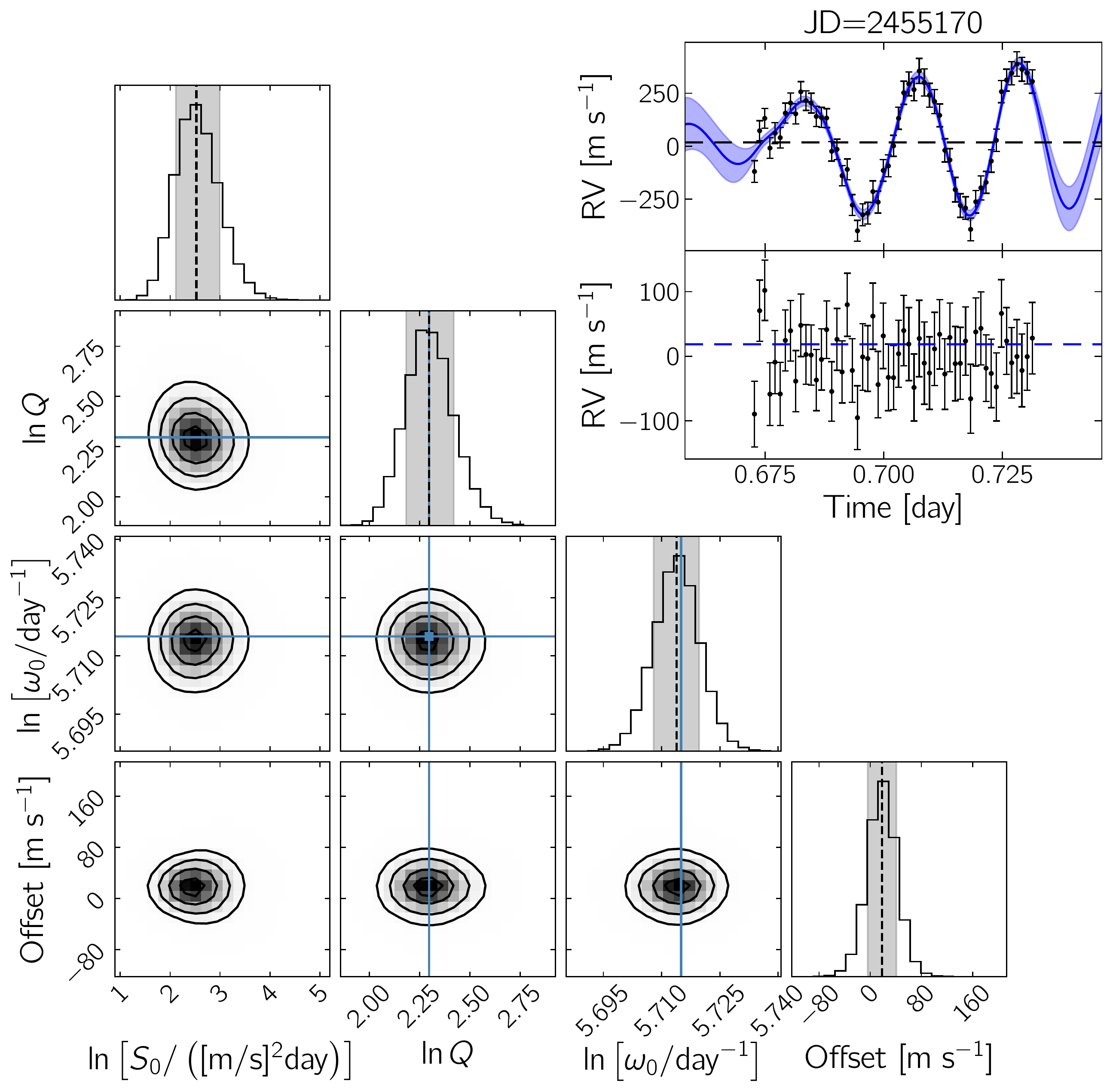}
    \caption{Example fit and posterior distribution for the SHO GP on RV Group \#15. The dashed lines and grey shaded areas show the median and the $1\sigma$ interval for each parameter (the three GP hyperparameters, and a constant offset). The solid blue lines show the median value of the training distribution for hyperparameters that were trained on the photometry. For $S_0$ and the RV offset, broad uniform priors were used. In the top right corner, the GP fit of the RV data in Group \#7 is shown as a solid blue line, with a shaded area showing the $1\sigma$ confidence interval. The bottom panel shows the residuals, the dashed blue line being the resulting offset of the group.}
    \label{fig:gp_rv_group}
\end{figure}

As mentioned in \S~\ref{sec:data_harps}, there is RV data available from 2003 to 2018. 
However, before 2008, the sampling is too sparse to exhibit the pulsations in the RV data. 
For this reason, \cite{lagrange_evidence_2019} used only the data from after 2008. We note that using a multi-sine parametric model fitted on the well-sampled post-2008 groups to fit groups from before 2008 is not possible because of temporal variations in phase and amplitude of the pulsation modes. A GP, on the other hand, does not retain any phase information and is not strongly affected by small changes in the amplitude of individual modes. Therefore, only the covariance properties of the pulsations matter for a given GP to effectively model various epochs. We use a GP to model all available RV data between 2003 and 2018. After splitting the data as described above, we obtain 36 groups: the same 28 post-2008 groups as \cite{lagrange_evidence_2019}, and 8 additional pre-2008 groups. We then fit each group with a GP model previously trained on the photometry.
To define the prior on the GP hyperparameters in the RV analysis, we use a kernel density estimate (KDE) of the posterior distributions from the photometry model (Figure~\ref{fig:photo_sho_post} for the SHO GP). We do not use such a prior for the offset and the amplitude, which are both expected to vary between the two datasets.

We first test the QP kernel, $k_\mathrm{QP}$, which properly accounts for the pulsations, but yields a large uncertainty on the RV offset. The uncertainties on RV offsets obtained with this kernel are approximately 4 times greater than those from the multi-sine parametric model used by \cite{lagrange_evidence_2019}.

The SHO kernel, on the other hand, results in offsets consistent with those from the multi-sine 
model, as shown in Figure~\ref{fig:gp_multisin_compare}.
It yields uncertainties about 1.5 times smaller, while requiring 3 hyperparameters instead of up to 90 parameters 
(a sum of up to 30 sine waves with 3 parameters each, \cite{lagrange_evidence_2019}). 
We also note that the RMS of the residuals from the SHO GP fit for all groups combined (with the offset subtracted from each group) is smaller by a factor of 1.4 than the RMS of the multi-sine residuals published by \cite{lagrange_evidence_2019}.
The SHO GP thus provides an effective alternative to the multi-sine model. 
The posterior distribution of the three GP hyperparameters and the constant offset, as well as the stellar activity fit for group \# 15 (December 4 2009), are shown in Figure~\ref{fig:gp_rv_group} as an example. 
Similar fits are obtained for the other groups and the hyperparameters are approximately constant 
throughout the groups, always staying within $1\sigma$ of other groups and of the training values. No long term drift is thus further introduced with the modelling. This is also true for the poorly sampled pre-2008 groups.

Both the QP and SHO GPs were a priori reasonable models for the pulsations of \Bpic{}. However, since the SHO GP is favoured by the BIC and also yields a better RV precision, we use RV offsets from this model in the remaining analysis.

% Placing this figure early so it renders on top of orbit fitting page
\begin{figure*}[ht!]
    \centering
    \includegraphics[width=0.95\linewidth]{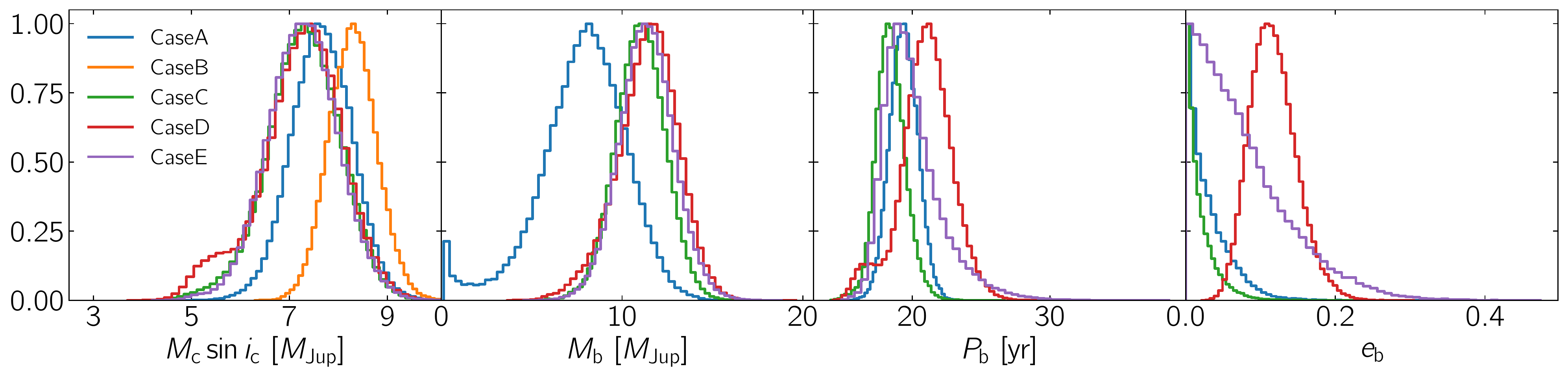}
    \caption{Posterior distribution for the minimum mass of \Bpic{} c, the mass, the period and the eccentricity of \Bpic{} b from the fit on RV data only. In Case B, only $M_\mathrm{c}\sin{i_\mathrm{c}}$ was fitted among these parameters. Fixing the orbit and mass of \Bpic{} b yields a greater, more precise mass for \Bpic{} c (Case B). Using post-2008 data only yields less precise constraints on the mass of \Bpic{} b with a small bimodality near $M_\mathrm{b} = 0$ \Mjup{}. Including pre-2008 data refines the constraints on the mass of \Bpic{} b, but its eccentricity and period are sensitive to the choice of priors. When a wide Gaussian prior is used on $P_b$ and a uniform prior is used on $e_b$ (Case E), low eccentricity and period are favoured.}
    \label{fig:rv_hists}
\end{figure*}

\section{Orbit Fitting}\label{sec:orbit}

The stellar activity signal has been effectively corrected from the RV, enabling to fit the two-planets from the RV offsets in order to constrain their orbital parameters and masses. In this section, we first explore the information content of the RV alone. Our goal is to assess whether our extended baseline and increased precision can lead to any constraint on the two planets. Then, we make use of all observables of the two planets in addition to the RVs, including relative astrometry from the ground using adaptive optics and interferometric data, in a joint fit. This complete analysis will provide robust constraints on the planet parameters.

\subsection{RV Offsets Only}\label{sec:rv_only}

We perform a two-planet fit of the RV offsets using the \texttt{radvel} Python package 
\citep{fulton_radvel:_2018}, along with \texttt{emcee} \citep{foreman-mackey_emcee:_2013} for 
MCMC sampling. To evaluate the impact of the extended RV baseline and the sensitivity to priors 
from relative astrometry, we explore five different Cases. In all Cases, we use uniform or log-uniform priors for the parameters of \Bpic{} c. For each planet, unless specified 
otherwise, we adjust five free parameters: the period $P$, the time of conjunction $T_\mathrm{c}$, 
$h = \sqrt{e}\cos{\omega}$, $k = \sqrt{e}\sin{\omega}$, and the RV semi-amplitude $K$. We also fit 
for a global RV offset $\gamma$ since the RVs were computed with respect to a reference spectrum 
\citep{galland_extrasolar_2005,lagrange_evidence_2019}. 
We use the alternative parameters $h$ and $k$ because they reduce 
the correlation between eccentricity $e$ and argument of periastron $\omega$\footnote{In 
\texttt{radvel}, $\omega$ is the argument of periastron of the star's orbit.} while also 
reducing the bias of the MCMC towards high eccentricities \citep{ford_improving_2006}. 
Additionally, we use the time of conjunction $T_\mathrm{c}$ instead of the time of periastron $T_\mathrm{p}$ 
since it is well defined for circular and low-eccentricity orbits. 

The five cases are the following:
\begin{itemize}
    \item Case A. We consider only the post-2008 RV offsets, obtained with the SHO GP. We impose Gaussian priors on the orbital parameters of \Bpic{} b based on the results from \cite{lagrange_post-conjunction_2019}. This is similar to the analysis from \cite{lagrange_evidence_2019}, but we use Bayesian priors instead of repeating the fit with fixed orbital parameters randomly drawn from the prior for \Bpic{} b.
    \item Case B. This is similar to Case A but we fix the orbital parameters of \Bpic{} b to the best-fit values from \cite{lagrange_post-conjunction_2019} and its mass to 10 \Mjup{}. As mentioned above, for each fit performed in \cite{lagrange_evidence_2019}, the orbital parameters of \Bpic{} are fixed to randomly drawn values from direct imaging results. $M_b$ was also held fixed, when using RV offsets only, and not all RV residuals. Case B is thus very close to the analysis from \citep{lagrange_evidence_2019} and enables a direct comparison of the parameters of \Bpic{} c between their analysis and ours.
    \item Case C. Here, we include all available RV offsets between 2003 and 2018, obtained with the SHO GP. The 8 additional RV offsets described in \S~\ref{sec:act_rv} provide an RV baseline of approximately 14 years. We use Gaussian priors from \cite{lagrange_post-conjunction_2019} on all parameters of \Bpic{} b except its period and its mass. For $M_b$, we use a wide Jeffreys priors, and for $P_b$, we apply a Gaussian prior centred at 22 yr with $\sigma = 4$ yr. This prior includes all period estimates published by~\cite{lagrange_post-conjunction_2019} ($\sim 20$ yr) and~\cite{nielsen_gemini_2020} ($\sim 24$ yr) within $1\sigma$.
    \item Case D. This is the same as Case C, but the priors from \cite{lagrange_post-conjunction_2019} are replaced with the slightly different priors from \cite{nielsen_gemini_2020} to test the sensitivity of the posteriors.
    \item Case E. To test whether the extended RV data can reveal a change in eccentricity as it did for the period, we use a different parametrization of the Keplerian. Adjusting the parameters $P$, $T_\mathrm{c}$, $e$, $\omega$ and $K$, which is already implemented in \texttt{radvel}, simplifies the use a uniform prior on $e_\mathrm{b}$ while imposing a Gaussian prior on $\omega$, whose value is consistent in both \cite{lagrange_post-conjunction_2019} and \cite{nielsen_gemini_2020}.
\end{itemize}

For each case, the adopted 
priors and the median of the parameter posterior distributions, along with the 1$\sigma$ 
uncertainties corresponding to the 16th and 84th percentiles, are summarized in 
Table~\ref{tab:rv_cases}. 
We also report the Bayesian Information Criterion (BIC) for each case to facilitate model comparison. Here, we provide an overview of the main results.

\begin{table*}
\movetabledown=50mm
\begin{rotatetable*}
    \caption{Two-Planet Fit With RV Data Only}\label{tab:rv_cases}
    \begin{center}
    \begin{tabular}{ccccccccccc} 
    \hline\hline             
    & \multicolumn{2}{c}{\multirow{2}{*}{Case A}} & \multicolumn{2}{c}{\multirow{2}{*}{Case B}} & \multicolumn{2}{c}{\multirow{2}{*}{Case C}} & \multicolumn{2}{c}{\multirow{2}{*}{Case D}} & \multicolumn{2}{c}{\textbf{Case E} } \\
    & & & & & & & & & \multicolumn{2}{c}{\textbf{(Adopted - RV Only)}} \\
    & \multicolumn{2}{c}{2008-2018} & \multicolumn{2}{c}{2008-2018} & \multicolumn{2}{c}{2003-2018} & \multicolumn{2}{c}{2003-2018} & \multicolumn{2}{c}{\textbf{2003-2018}} \\
    \cmidrule(lr){2-3} \cmidrule(lr){4-5} \cmidrule(lr){6-7} \cmidrule(lr){8-9} \cmidrule(lr){10-11}
    & Prior & Posterior & Prior & Posterior & Prior & Posterior & Prior & Posterior & \textbf{Prior} & \textbf{Posterior} \\
    \hline
    $P_\text{b}$ [yr]\tablenotemark{a}
        & $\mathcal{N}(20.3, 1.1)$ & $19.4 \pm 1.0$
        & \multicolumn{2}{c}{Fixed to 20.3}
        & $\mathcal{N}(22, 4)$ & $18.4^{+1.1}_{-1.0}$
        & See C & $21.1 \pm 1.8$
        & See C & $\mathbf{19.6^{+2.4}_{-1.5}}$\\
    $T_{\mathrm{c},\text{b}}$ [JD-2 450 000]
        & $\mathcal{N}(8010, 15)$ & $8012\pm 15$
        & \multicolumn{2}{c}{Fixed to 8010}
        & See A & $8013 \pm 15$
        & See A & $8013 \pm 15$
        & \textbf{See A} & $\mathbf{8012 \pm 15}$\\
    $h_{\text{b}} = \sqrt{e_{\text{b}}}\cos{\omega_{\text{b}}}$\tablenotemark{b}
        & $\mathcal{N}(0.09, 0.10)$ & $0.14 \pm 0.10$
        & \multicolumn{2}{c}{Fixed to 0.09}
        & See A & $0.10 \pm 0.09$
        & $\mathcal{N}(0.33, 0.05)$ & $0.32 \pm 0.05$
        & \textbf{---} & \textbf{---} \\
    $k_{\text{b}} = \sqrt{e_{\text{b}}}\sin{\omega_{\text{b}}}$\tablenotemark{b}
        & $\mathcal{N}(0.03, 0.04)$ & $0.03 \pm 0.04$
        & \multicolumn{2}{c}{Fixed to 0.03}
        & See A & $0.03 \pm 0.04$
        & $\mathcal{N}(0.11, 0.03)$ & $0.10 \pm 0.3$
        & \textbf{---} & \textbf{---} \\
    $K_{\text{b}}$ [m s$^{-1}$]
        & $\mathcal{J}^{*}(1, 200)$  & $58^{+15}_{-17}$
        & \multicolumn{2}{c}{Fixed to 69.4}
        & See A & $81 \pm 10$
        & See A & $81^{+10}_{-11}$
        & \textbf{See A} & $\mathbf{82 \pm 10}$\\
    $e_\text{b}$
        & --- & $0.021^{+0.035}_{-0.017}$
        & \multicolumn{2}{c}{Fixed to 0.010}
        & --- & $0.013^{+0.023}_{-0.010}$
        & --- & $0.115^{+0.033}_{-0.028}$
        & $\bm{\mathcal{U}}\mathbf{(0.0, 0.5)}$ & $\mathbf{0.062^{+0.075}_{-0.044}}$ \\
    $\omega_\text{b}$ [deg]
        & --- & $13^{+31}_{-15}$
        & \multicolumn{2}{c}{Fixed to 20}
        & --- & $16^{+49}_{-21}$
        & --- & $17^{+6}_{-5}$
        & $\bm{\mathcal{N}}\mathbf{(20, 10)}$ & $\mathbf{16 \pm 11}$\\
    $a_\text{b}$ (AU)
        & --- & $8.7 \pm 0.3$
        & \multicolumn{2}{c}{Fixed to 9.1}
        & --- & $8.4^{+0.4}_{-0.3}$
        & --- & $9.2 \pm 0.5$
        & \textbf{---} & $\mathbf{8.8^{+0.7}_{-0.5}}$\\
    $M_\text{b}$ [M$_{\text{Jup}}$)
        & --- & $8.0^{+2.1}_{-2.4}$
        & \multicolumn{2}{c}{Fixed to 10.0}
        & --- & $11.0 \pm 1.4$
        & --- & $11.5^{+1.6}_{-1.8}$
        & \textbf{---} & $\mathbf{11.4 \pm 1.5}$\\
    $P_\text{c}$ [yr]\tablenotemark{a}
        & $\mathcal{U}(2.74, 3.83)$ & $3.29 \pm 0.03$
        & See A & $3.32 \pm 0.03$
        & See A & $3.32^{+0.03}_{-0.04}$
        & See A & $3.32^{+0.04}_{-0.05}$
        & \textbf{See A} & $\mathbf{3.32 \pm 0.04}$\\  
    $T_{\mathrm{c},\text{c}}$ [JD-2 450 000]
        & $\mathcal{U}(7500,9000)$ & $8118^{+32}_{-33}$
        & See A & $8152^{+23}_{-20}$
        & See A & $8144^{+29}_{-28}$
        & See A & $8141^{+31}_{-35}$
        & \textbf{See A} & $\mathbf{8143^{+31}_{-28}}$\\
    $h_{\text{c}} = \sqrt{e_{\text{c}}}\cos{\omega_{\text{c}}}$\tablenotemark{b}
        & $\mathcal{U}(-1,1)$ & $0.11^{+0.18}_{-0.17}$
        & See A & $0.04 \pm 0.12$
        & See A & $0.03 \pm 0.14$
        & See A & $0.04^{+0.16}_{-0.15}$
        & \textbf{---} & \textbf{---} \\
    $k_{\text{c}} = \sqrt{e_{\text{c}}}\sin{\omega_{\text{c}}}$\tablenotemark{b}
        & $\mathcal{U}(-1,1)$ & $-0.23^{+0.27}_{-0.16}$
        & See A & $-0.38^{+0.10}_{-0.08}$
        & See A & $-0.34^{+0.19}_{-0.12}$
        & See A & $-0.32^{+0.28}_{-0.13}$
        & \textbf{---} & \textbf{---} \\
    $K_{\text{c}}$ [m s$^{-1}$]
        & $\mathcal{J}^{*}(1, 200)$ & $101 \pm 8$
        & See A & $110 \pm 7$
        & See A & $96 \pm 10$
        & See A & $96^{+10}_{-11}$
        & \textbf{See A} & $\mathbf{96 \pm 10}$\\
    $e_\text{c}$
        & --- & $0.111^{+0.082}_{-0.061}$
        & --- & $0.161^{+0.064}_{-0.057}$
        & --- & $0.141^{+0.084}_{-0.076}$
        & --- & $0.138^{0.102}_{-0.081}$
        & $\bm{\mathcal{U}}\mathbf{(0, 0.99)}$ & $\mathbf{0.130^{+0.083}_{-0.080}}$ \\
    $\omega_\text{c}$ [deg]
        & --- & $-64^{+75}_{-34}$
        & --- & $-84^{+20}_{-17}$
        & --- & $-84^{+34}_{-23}$
        & --- & $-81^{+67}_{-25}$
        & $\bm{\mathcal{U}}\mathbf{(-200, 160)}$ & $\mathbf{-84^{+38}_{-25}}$\\
    $a_\text{c}$ [AU]
        & --- & $2.68 \pm 0.03$
        & --- & $2.69 \pm 0.02$
        & --- & $2.69 \pm 0.03$
        & --- & $2.69 \pm 0.03$
        & \textbf{---} & $\mathbf{2.69 \pm 0.03}$\\
    $M_\mathrm{c}\sin{i_\mathrm{c}}$ [M$_{\mathrm{Jup}}$]
        & --- & $7.7 \pm 0.6$
        & --- & $8.3 \pm 0.5$
        & --- & $7.3 \pm 0.8$
        & --- & $7.3^{+0.8}_{-0.9}$
        & \textbf{---} & $\mathbf{7.3 \pm 0.8}$\\
    $\gamma$ [m s$^{-1}$]
        & $\mathcal{U}(-100,100)$ & $-21 \pm 9$
        & See A & $-31 \pm 5$
        & See A & $-26 \pm 9$
        & See A & $-33^{+12}_{-11}$
        & \textbf{See A} & $\mathbf{-30 \pm 10}$\\ \hline
    BIC\tablenotemark{c}
        & \multicolumn{2}{c}{383}
        & \multicolumn{2}{c}{367}
        & \multicolumn{2}{c}{496}
        & \multicolumn{2}{c}{500}
        & \multicolumn{2}{c}{$\mathbf{498}$}\\
    \hline
    \end{tabular}
    \end{center}
    \tablecomments{$\mathcal{J}(a,b)$ denotes a Jeffrey's prior between $a$ and $b$, $\mathcal{N}(\mu, \sigma)$ 
    denotes a Gaussian prior, $\mathcal{J}^{*}(k,m)$ denotes a modified Jeffrey's prior that behaves as a uniform 
    below $k$ and as a Jeffrey's prior between $k$ and $m$.}
    \tablenotetext{a}{Periods are given in years for convenience, but the units used for the analysis were days.}
    \tablenotetext{b}{When fitting $h_j$ and $k_j$, we also always ensure that $e_j = h_j^2+k_j^2 < 1$.}
    \tablenotetext{c}{BIC values can only be compared when the two model fit the same 
    values of independent variable, i.e. Case A and B should be compared separately from Cases C, D and E.}
\end{rotatetable*}
\end{table*}

As mentioned above, \cite{lagrange_evidence_2019} used all of the RV residuals for each group between 2008 and 2018 to constrain the mass of \Bpic{} b. When only the multi-sine offsets are used, the mass of \Bpic{} b is not constrained and has to be held fixed. In Case A, thanks to the  increased precision provided by the SHO GP, we constrain the orbital parameters and masses of both planets using the 28 RV 
offsets only. We find $M_\mathrm{b} = 8.0^{+2.1}_{-2.4}$ \Mjup{} and $M_\mathrm{c} = 7.7 \pm 0.6$ \Mjup{}. These masses  are both lower than those found by \cite{lagrange_evidence_2019}. 
As shown in Figure~\ref{fig:rv_hists}, the posterior distribution of $M_\mathrm{b}$ includes a small bimodality near 0 \Mjup{}.
The results from this fit should therefore be interpreted carefully. 

In Case B, we obtain  results consistent with those of \cite{lagrange_evidence_2019} and \cite{nielsen_gemini_2020} for  the orbital parameters and the mass of \Bpic{} c. 
The main difference is in eccentricity, with $e = 0.161^{+0.064}_{-0.057}$ being lower 
than values from the literature, but still consistent within $2\sigma$. This result suggests that 
the GP RV offsets might favor slightly lower eccentricities than multi-sine offsets from 
\cite{lagrange_evidence_2019}. The BIC for Cases A and B 
yield $\Delta \text{BIC}_{A-B}$ = 16, providing very strong evidence in favour of Case B. This is 
mostly attributable to the difference in the number of parameters. The fact that allowing more 
free parameters does not yield better agreement with the data is another sign that 
parameters of \Bpic{} b are not totally well-constrained by the 2008-2018 GP offsets only, and 
that results from Case A should be interpreted with caution. However, the fact that they enable 
constraints on the signal of \Bpic{} b with RV offsets only does show that the smaller error bars of the GP 
offsets provide more information on the planetary orbits compared to the multi-sine offsets.

In Cases C and D, we first tried using uniform priors for the orbit of \Bpic{} b, but several parameters were 
not constrained properly. Hence, as described above, we use Gaussian priors for all parameters of 
\Bpic{} b except its mass. Performing a two-planet fit, we find $P_\mathrm{b} = 18.4^{+1.1}_{-1.0}$ yr 
for Case C and $P_\mathrm{b} = 21.1 \pm 1.8$ yr for Case D. Both Cases seem to favour lower 
periods than references from which the priors were taken, but are consistent with their 
respective reference within $2\sigma$. As shown in Figure \ref{fig:rv_hists}, the eccentricity 
is also sensitive to the choice of priors with Case C allowing circular orbits and Case D 
favouring slightly eccentric orbits. The masses, on the other hand, show low sensitivity to the 
choice of priors. 
Case C yields $\left[M_\mathrm{b}, M_\mathrm{c}\right] = \left[11.0 \pm 1.4, 7.3 \pm 0.8\right]$ \Mjup{} 
while Case D gives $\left[M_\mathrm{b}, M_\mathrm{c}\right] = \left[11.5^{+1.6}_{-1.8}, 7.3^{+0.8}_{-0.9}\right]$ \Mjup{}.  
We note that the BIC values shown in Table~\ref{tab:rv_cases} provides evidence in favour of Case C with priors from \cite{lagrange_post-conjunction_2019}.

\begin{figure}
    \centering
    \includegraphics[width=\linewidth]{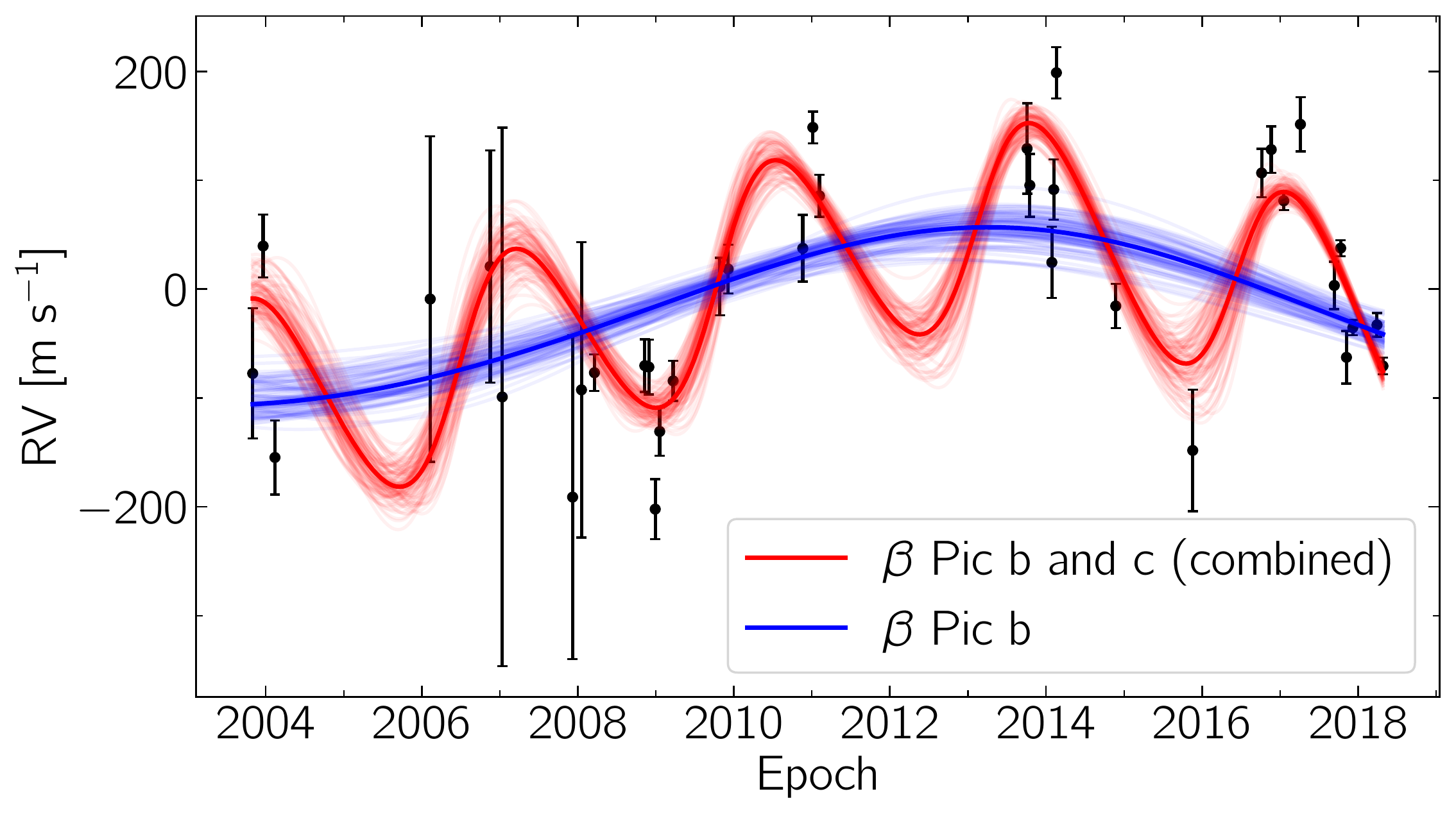}
    \caption{Example fit using RV data only. All RV offsets between 2003 and 2018 are shown in black. The fit of \Bpic{} b and c combined is shown as a solid red line, and the corresponding solution for \Bpic{} b only is shown in blue. Both are shown along with 100 random MCMC samples. This fit corresponds to Case E in Table~\ref{tab:rv_cases}}
    \label{fig:rv_fit_e}
\end{figure}

\begin{figure*}
    \centering
    \includegraphics[width=0.95\linewidth]{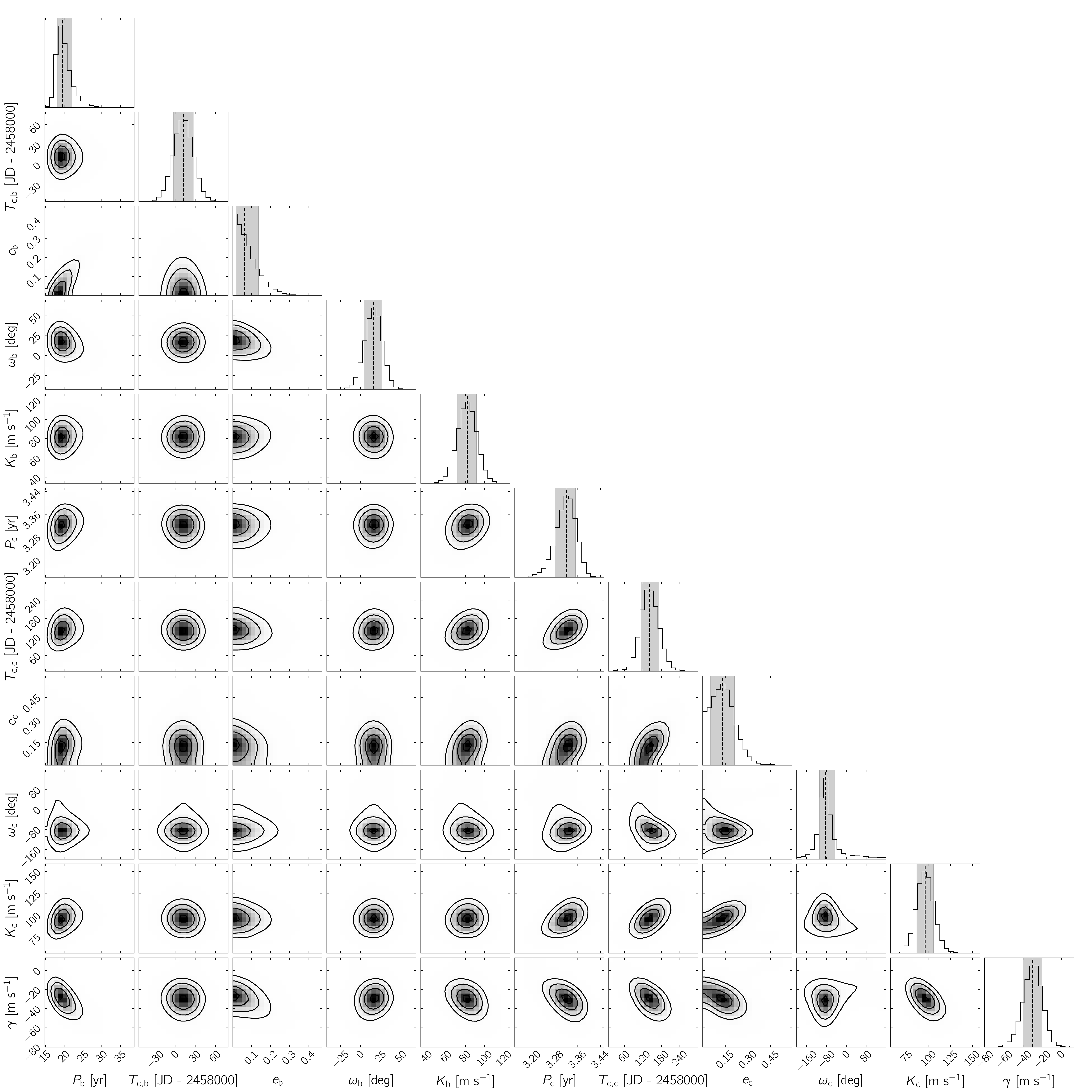}
    \caption{Posterior distribution of the Case E RV fit. The dashed lines and grey shaded areas show the median and the $1\sigma$ interval for each parameter. Low-eccentricity orbits are preferred for \Bpic{} b. This corresponds to shorter orbital periods due to the correlation between these two parameters.}
    \label{fig:corner_e}
\end{figure*}

In Case E, we find $P_\mathrm{b} = 19.6^{+2.4}_{-1.5}$ yr and $e_\mathrm{b} = 0.062^{+0.072}_{-0.045}$. 
Therefore, the RV data favours low period and circular orbits when these 
two parameters are not constrained ($e_b$) or weakly constrained ($P_b$) with priors from direct imaging. Constraints on both masses remain unaffected by this change 
of parametrization and priors (see Figure~\ref{fig:rv_hists}). Figure~\ref{fig:rv_fit_e} shows 
the RV fit corresponding to Case E and Figure~\ref{fig:corner_e} shows the posterior distribution 
of the fitted parameters. The distributions are generally well-constrained. The apparent 
correlation between $e_\mathrm{b}$ and $P_\mathrm{b}$ is consistent with results from the literature where longer 
periods are generally associated with more eccentric orbits \citep{lagrange_post-conjunction_2019,nielsen_gemini_2020,dupuy_model-independent_2019}. We note that the BIC favours Case E to Case D, but not Case C. In both cases, the evidence is only marginal with $\Delta\mathrm{BIC}=2$. This suggests that priors from astrometry can still 
provide valuable information about the orbit of \Bpic{} b despite the better constraints provided by the GP offsets and the extended RV baseline.

Throughout the 5 Cases, \Bpic{} c is always detected with a false-alarm probability $\mathrm{FAP}<2.2 \times 10^{-6}$, and its orbital parameters and its mass show very low sensitivity to 
the choice of orbital priors for \Bpic{} b. The constraints on $M_\mathrm{b}$ are also unaffected by changes 
in these priors. As mentioned above, the improved precision of GP offset plays an important role 
in our ability to constrain $M_\mathrm{b}$ in Case A. However, one can clearly see in Figure~\ref{fig:rv_hists} that the additional 2003-2008 RV coverage improves the mass 
estimate, yielding $M_\mathrm{b}= 11.4 \pm 1.5$ \Mjup{} (Case E) and completely ruling out solutions where b 
is undetected. This mass is consistent with the indicative mass of $M_\mathrm{b} \sim 10$ 
\Mjup{} reported by \cite{lagrange_evidence_2019}. Thanks to 
the extended RV baseline, we are now able to constrain the mass of both planets from RV 
data only with very good precision and low sensitivity to the choice of priors.

\subsection{RV Offsets and Relative Astrometry}\label{sec:joint_rv_astro}

\begin{table*}
    \caption{Two-Planet Joint Fit of RV and Relative Astrometry}\label{tab:joint}
    \begin{center}
    \begin{tabular}{ccccc} 
        \hline\hline             
        & \multicolumn{2}{c}{\multirow{2}{*}{Joint Fit}} 
        & \multicolumn{2}{c}{\textbf{Joint Fit With Gravity}} \\
        & & & \multicolumn{2}{c}{\textbf{(Adopted - Final)}}\\
        \cmidrule(lr){2-3} \cmidrule(lr){4-5}
        & Prior & Posterior & \textbf{Prior} & \textbf{Posterior} \\
        \hline
        $P_\text{b}$ [yr]
            & --- & $21.2^{+2.4}_{-0.9}$
            & \textbf{---} & $\mathbf{21.1^{+2.0}_{-0.8}}$\\
        $T_{\mathrm{c},\text{b}}$ [JD-2 450 000]
            & $\mathcal{U}(7000,9000)$ & $8012.2 \pm 2.3$
            & $\bm{\mathcal{U}}\mathbf{(7000,9000)}$ & $\mathbf{8011.2 \pm 0.5}$\\
        $h_{\text{b}} = \sqrt{e_{\text{b}}}\cos{\omega_{\text{b}}}$\tablenotemark{a}
            & $\mathcal{U}(-1, 1)$ & $0.156^{+0.137}_{-0.175}$
            & $\bm{\mathcal{U}}\mathbf{(-1, 1)}$ & $\mathbf{0.146^{+0.123}_{-0.160}}$\\
        $k_{\text{b}} = \sqrt{e_{\text{b}}}\sin{\omega_{\text{b}}}$\tablenotemark{a}
            & $\mathcal{U}(-1, 1)$ & $0.074^{+0.061}_{-0.101}$
            & $\bm{\mathcal{U}}\mathbf{(-1, 1)}$ & $\mathbf{0.077^{+0.058}_{-0.098}}$\\
        $e_\text{b}$
            & --- & $0.033^{+0.069}_{-0.027}$
            & \textbf{---} & $\mathbf{0.029^{+0.061}_{-0.024}}$\\
        $\omega_\text{b}$ [deg]
            & --- & $22^{+18}_{-67}$
            & \textbf{---} & $\mathbf{25^{+15}_{-62}}$\\
        $a_\text{b}$ [AU]
            & $\mathcal{J}(1,100)$ & $9.3^{+0.7}_{-0.3}$
            & $\bm{\mathcal{J}}\mathbf{(1,100)}$ & $\mathbf{9.2^{+0.6}_{-0.2}}$\\
        $M_\text{b}$ [M$_{\text{Jup}}$]
            & $\mathcal{U}(1, 20)$ & $11.8 \pm 1.4$
            & $\bm{\mathcal{U}}\mathbf{(1, 20)}$ & $\mathbf{11.7 \pm 1.4}$\\
        $\Omega_\text{b}$ [deg]
            & $\mathcal{U}(18, 90)$ & $32.07 \pm 0.08$
            & $\bm{\mathcal{U}}\mathbf{(18, 90)}$ & $\mathbf{32.00^{+0.06}_{-0.05}}$\\
        $i_\text{b}$ [deg]
            & $\sin{i}$ & $88.97 \pm 0.09$
            & $\bm{\sin{i}}$ & $\mathbf{88.88^{+0.04}_{-0.03}}$\\
        $P_\text{c}$ [yr]\tablenotemark{a}
            & --- & $3.36 \pm 0.03$
            & \textbf{---} & $\mathbf{3.36 \pm 0.03}$\\
        $T_{\mathrm{c},\text{c}}$ [JD-2 450 000]
            & $\mathcal{U}(7500,9000)$ & $8177^{+31}_{-27}$
            & $\bm{\mathcal{U}}\mathbf{(7500,9000)}$ & $\mathbf{8177^{+31}_{-27}}$\\
        $h_{\text{c}} = \sqrt{e_{\text{c}}}\cos{\omega_{\text{c}}}$\tablenotemark{a}
            & $\mathcal{U}(-1,1)$ & $0.024^{+0.102}_{-0.103}$
            & $\bm{\mathcal{U}}\mathbf{(-1,1)}$ & $\mathbf{0.025 \pm 103}$\\
        $k_{\text{c}} = \sqrt{e_{\text{c}}}\sin{\omega_{\text{c}}}$\tablenotemark{a}
            & $\mathcal{U}(-1,1)$ & $-0.442^{+0.085}_{-0.080}$
            & $\bm{\mathcal{U}}\mathbf{(-1,1)}$ & $\mathbf{-0.442^{+0.086}_{-0.081}}$\\
        $e_\text{c}$
            & --- & $0.206^{+0.074}_{-0.063}$
            & \textbf{---} & $\mathbf{0.206^{+0.074}_{-0.063}}$\\
        $\omega_\text{c}$ [deg]
            & --- & $-87^{+14}_{-13}$
            & \textbf{---} & $\mathbf{-87^{+14}_{-13}}$\\
        $a_\text{c}$ [AU]
            & $\mathcal{J}(2,3.5)$ & $2.71 \pm 0.02$
            & $\bm{\mathcal{J}}\mathbf{(2,3.5)}$ & $\mathbf{2.71 \pm 0.02}$\\
        $M_\mathrm{c}\sin{i_\mathrm{c}}$ [M$_{\mathrm{Jup}}$]
            & $\mathcal{U}(1, 20)$ & $8.5 \pm 0.5$
            & $\bm{\mathcal{U}(1, 20)}$ & $\mathbf{8.5 \pm 0.5}$\\
        $\Mtot{}$ [$M_\odot$]
            & $\mathcal{U}(1.4, 2.0)$ & $1.76 \pm 0.03$
            & $\bm{\mathcal{U}}\mathbf{(1.4, 2.0)}$ & $\mathbf{1.76 \pm 0.03}$\\
        $\pi$ [mas]
            & $\mathcal{N}(51.44, 0.12)$ & $51.44 \pm 0.12$
            & $\bm{\mathcal{N}}\mathbf{(51.44, 0.12)}$ & $\mathbf{51.44 \pm 0.12}$\\
        $\rho_\mathrm{S}/\rho_\mathrm{G}$
            & $\mathcal{U}(0.5, 1.5)$ & $0.99 \pm 0.6$
            & $\bm{\mathcal{U}}\mathbf{(0.5, 1.5)}$ & $\mathbf{0.99 \pm 0.6}$\\
        $\theta_\mathrm{S} - \theta_\mathrm{G}$ [deg]
            & $\mathcal{N}(0.0, 0.5)$ & $-0.14 \pm 0.48$
            & $\bm{\mathcal{N}}\mathbf{(0.0, 0.5)}$ & $\mathbf{-0.15 \pm 0.48}$\\
        $\gamma$ [m s$^{-1}$]
            & $\mathcal{U}(-100,100)$ & $-42^{+7}_{-8}$
            & $\bm{\mathcal{U}}\mathbf{(-100,100)}$ & $\mathbf{-42 \pm 7}$\\
        \hline
        \end{tabular}
        \tablenotetext{a}{When fitting $h_j$ and $k_j$, we also always ensure that $e_j = h_j^2+k_j^2 < 1$.}
        \end{center}
\end{table*}

To further assess the impact of the extended RV baseline and of the GP offsets, and to better 
characterize the system as whole, we perform a joint fit of all available RV and relative 
astrometry data. To do so, we still use \texttt{radvel}, but we include custom models and  
likelihoods to account for relative astrometry. We still rely 
on on the Keplerian solver from \texttt{radvel}, but we then derive values of Separation (Sep.) 
and position angle (PA) using a model similar to the one used in \texttt{orbitize!} 
\citep{blunt_orbitize_2020}. 

In this joint model, we fit the following orbital parameters for both planets: semimajor axis $a$, 
time of conjunction $T_\mathrm{c}$, $h = \sqrt{e}\cos{\omega}$ and $k = \sqrt{e}\sin{\omega}$. To account 
for the relative astrometry of \Bpic{} b, we also adjust its inclination $i_\mathrm{b}$, the position 
angle of ascending node $\Omega_\mathrm{b}$, the parallax $\pi$ and the total mass \Mtot{}. Since the 
inclination of \Bpic{} b is constrained by relative astrometry, we fit for its mass, $M_\mathrm{b}$, 
directly. This is not the case of \Bpic{} c so we use $M_\mathrm{c}\sin{i_\mathrm{c}}$. 
As in \S~\ref{sec:rv_only}, we also include a global RV offset $\gamma$. 
Finally, as pointed out by \cite{nielsen_gemini_2020}, there is evidence for a systematic position angle offset between GPI and SPHERE. 
Following their analysis, we include two additional free parameters: a multiplicative 
correction, $\rho_S/\rho_G$, in separation and an additive offset position angle, $\theta_S - 
\theta_G$, both applied to GPI data points. This adds up to a total of 17 free parameters. We use 
uniform or log-uniform priors on all parameters except the parallax for which we use a Gaussian prior 
corresponding to $\pi = 51.44 \pm 12$ mas, as measured by \textit{Hipparcos} 
\citep{van_leeuwen_validation_2007}. 
For $\Omega_\mathrm{b}$, we use a uniform prior between 
$\pi/10$ and $\pi/2$, to account for the planetary RV measurement by \cite{snellen_fast_2014}. For the inclination, we use a prior uniform in $\sin(i_b)$. 
A detailed list of adopted priors is provided in Table~\ref{tab:joint}. We perform two joint 
fits: one with and one without the VLTI/GRAVITY data point from September 2018. The posterior 
distribution of all parameters for both fits is shown in Appendix \ref{app:all_post}.

\begin{figure*}
    \centering
    \includegraphics[width=0.95\linewidth]{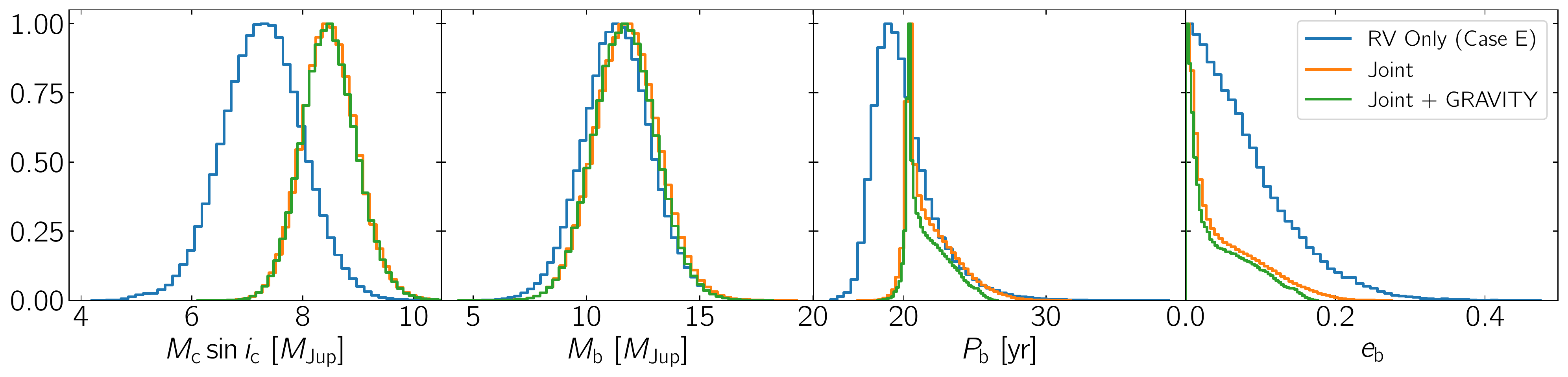}
    \caption{Posterior distribution for the minimum mass of \Bpic{} c, the mass, the period and the eccentricity of \Bpic{} b. The parameter distributions are shown for a fit of RV data only (Case E), a joint fit of RV data and relative astrometry, and the same joint fit, but including the high-precision measurement from VLTI/GRAVITY. The joint fit generally provides better precision than the RV-only fit. Low-eccentricity and short period orbits are also more strongly favoured. The inclusion of the GRAVITY measurement slightly improves the precision of most parameters, but does not cause a dramatic change in any of them.}
    \label{fig:joint_hists}
\end{figure*}

We perform a first fit using all available data except the GRAVITY measurement from 2018. 
In general, values are consistent with the RV-only fit, but have better precision. While 
constraints on the mass of \Bpic{} b barely change with the inclusion of relative astrometry, 
yielding $M_\mathrm{b} = 11.8 \pm 1.4$ \Mjup{}, the mass of \Bpic{} c is now measured at 
$M_\mathrm{c} = 8.5 \pm 0.5$ \Mjup{}, an increase of about 1 \Mjup{} compared to the RV fits of \S~\ref{sec:rv_only}. 
This constraint on $M_\mathrm{c}$ is consistent with the literature 
\citep{lagrange_evidence_2019,nielsen_gemini_2020}. However, accounting for \Bpic{} c does not 
cause a decrease in $M_\mathrm{b}$ as reported by \cite{nielsen_gemini_2020} ($\sim 8$ \Mjup{}) and, to a 
lesser extent, \cite{lagrange_evidence_2019} ($\sim 10$ \Mjup{}). 
This is likely due to the extended RV baseline, to the increased 
precision of the GP offsets compared to the multi-sine offsets, to the fact that we do not 
include absolute astrometry, or to a combination of those 3 factors. The tendency 
towards short-period (20-21 yr), low-eccentricity orbits observed when fitting RV data only is 
confirmed in the joint fit, and both parameters are constrained more precisely, as shown in Figure~\ref{fig:joint_hists}. 
We note that the measured eccentricity of \Bpic{} c is now slightly higher 
at $e_\mathrm{c} = 0.206^{+0.074}_{-0.063}$ and is in better agreement with the literature 
\citep{lagrange_evidence_2019,nielsen_gemini_2020}. Similarly to \cite{nielsen_gemini_2020}, we 
find no evidence of a systematic offset $\rho_G/\rho_S$ in separation. In PA, however, the correction stays 
unconstrained after the MCMC has reached convergence. In a attempt to better constrain this 
parameter, we use a wide Gaussian prior centred at $0^\circ$ with $\sigma = 0.5^\circ$. This prior includes the correction of $-0.47^\circ \pm 0.14^\circ$ found by \citet{nielsen_gemini_2020} within $1\sigma$. We then find a correction $\theta_G - \theta_S = -0.14^\circ \pm 0.48^\circ$. This corresponds to a 
true north offset as reported by \cite{nielsen_gemini_2020}, but the offset is smaller, and is consistent with $0^\circ$ well within $1\sigma$. In addition, we performed a fit without including any correction and found that the posterior distribution of orbital parameters was unaffected.

\begin{figure*}
    \centering
    \includegraphics[width=0.95\linewidth]{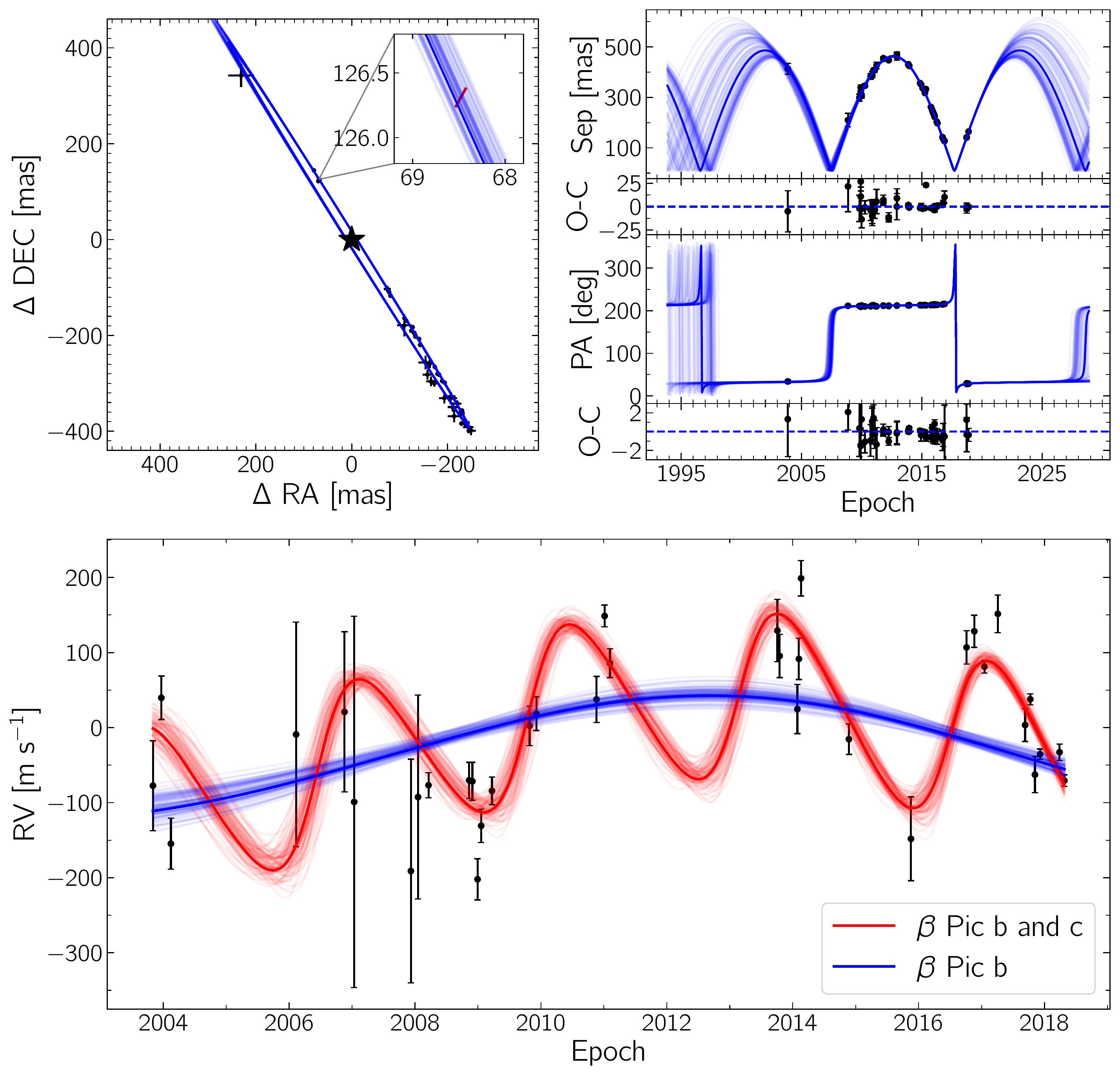}
    \caption{Joint fit of RV offsets with all available relative astrometry of \Bpic{} b. In all panels, the orbit of \Bpic{} b is shown in blue. In the top left panel, it is shown along with all relative astrometry measurements. The VLTI/GRAVITY measurement appears in an inset to show how its high precision constraints the fit. All other relative astrometry measurements were fit in separation and position angle rather than RA and DEC, as shown in the top right panel. The bottom panel shows the 36 RV offsets along with the joint signal of both planets in addition to the signal or \Bpic{} b alone. All fits are shown with 100 random samples from the MCMC.}
    \label{fig:joint_fit}
\end{figure*}

When including the GRAVITY data point in our fit, we must add a second custom term to the log-likelihood since 
the two coordinates (\dra{} and \ddec{}) are correlated. 
To do so, we first calculate the Pearson correlation coefficient corresponding to the covariance 
matrix of Equation \ref{eq:gravity}. This allows us to obtain the probability density function (PDF) 
of a bivariate normal distribution centred on the model values, but with an analytical solution that does not require any matrix inversion, taking advantage of the pre-computed correlation coefficient. 
We then evaluate this PDF at the coordinates of the data point to obtain the 
likelihood before taking its logarithm and adding it to the other two 
likelihoods (RV and other relative astrometry). However, the covariance matrix $\Sigma$ from 
Equation \ref{eq:gravity}, reported by \cite{nowak_peering_2020}, 
is not positive semi-definite. This yields an invalid Pearson correlation coefficient $r < -1$. 
This issue is likely due to an approximation in the reported covariance matrix. 
We therefore simply set the off-axis elements of the covariance matrix to 
$\sigma_{\dra{},\ddec{}} = \sigma_{\ddec{},\dra{}} = -0.00348$ mas$^2$. 
This small change of $0.00002$ mas$^2$ reestablishes the positive 
semi-definiteness of $\Sigma$ and puts the Pearson coefficient back in the expected interval 
$-1 < r < 1$\footnote{Note that the original covariance matrix does not perfectly match the error bars shown in Figure 2 of \citet{nowak_peering_2020} since both were made separately at different moments of the analysis and not updated for publication (J. Wang, private communication)}. The principal components of this adjusted covariance matrix 
give the error bars on the GRAVITY measurement, as shown in the top left panel 
Figure~\ref{fig:joint_fit}. 
Once this fix is applied, we can fit for all available relative astrometry, including the GRAVITY measurement.

As reported in Table~\ref{tab:joint}, the inclusion of the GRAVITY measurement yields results 
consistent with the joint fit performed without including it. However, in general, the precision 
is slightly improved when including the \mbox{GRAVITY} measurement, which is more precise than other 
relative astrometric measurements by more than an order of magnitude. This precision improvement is 
marginal and can thus hardly be seen by looking at the posterior distribution of parameters, but a 
careful inspection of Figure~\ref{fig:joint_hists} does reveal a sharper peak in $P_\mathrm{b}$ and a 
slightly stronger preference for low-eccentricity orbits. Even when using relative 
astrometry only, \cite{nowak_peering_2020} reported a preference for slightly eccentric orbits 
($e_\mathrm{b} = 0.15^{+0.05}_{-0.04}$), which we do not observe here. This could be because they do not account for \Bpic{} c in their analysis while we account for both planets using the RV offsets (see \cite{hara_bias_2019} for a detailed discussion of such possible eccentricity biases).
Figure~\ref{fig:joint_fit} shows the joint fit of RV data with all available relative astrometry, including the GRAVITY measurement. 
The fit is in good agreement with the high-precision GRAVITY measurement, as shown in the top left 
panel.

\section{Discussion}\label{sec:discussion}

The orbital properties and masses of \Bpic{} b and c provide additional knowledge about the full planetary system architecture, including the debris disk, and its evolution with time. We discuss the impact of our findings on these two aspects and propose future directions to further improve the system characterization.

Our orbital parameters for \Bpic{} b confirm that the planet is misaligned with the main disk and consistent with the inner warp \citet{lagrange_constraints_2012}. A direct detection of \Bpic{} c would also provide a measurement of the inclination of its orbit, further revealing the inner architecture of the system with respect to the disk. We show the orbit of \Bpic{} b to have a very small eccentricity ($<0.06$). This value is still consistent 
at $1\sigma$ with the planet being responsible for the exocomets ($e>0.05$) \citep{beust_mean-motion_1996}, but the inclusion of the more eccentric orbit of \Bpic{} c should be investigated.

The mass of directly-imaged planets cannot be obtained from direct spectrophotometric measurements. Without complementary data like RV or absolute astrometry, the estimate of this parameter solely relies on theoretical models, which predict the temporal evolution of the luminosity of a planet for a variety of masses. The ages of directly imaged planet hosts is usually well constrained \citep[e.g.][]{mamajek_age_2014}; the bolometric luminosity of the planets are computed from their spectral energy distributions. However, the evolutionary models are not calibrated at young ages ($<1$Gyr) and low masses ($<72$\Mjup). The model-based masses of the directly imaged planets might then be incorrect. The independent mass measurement of \Bpic{} b, derived from the RV data, offers a rare opportunity to be compared against the predictions of the evolutionary models. In the following, we will compute the predicted mass according to the known age and luminosity of \Bpic{} b and see whether the modelled-derived age is consistent with our dynamical mass.

\begin{figure*}
    \centering
    \includegraphics[width=\linewidth]{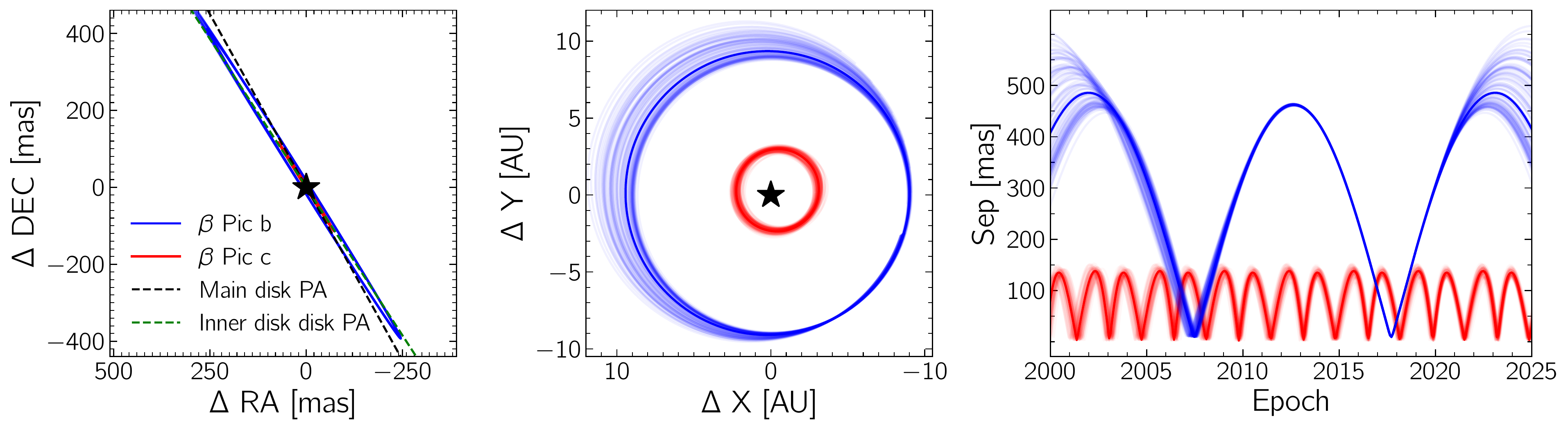}
    \caption{Orbit solutions showing the system architecture in various coordinates. The left panel shows the orbit of both planets in RA and DEC, assuming coplanar orbits for both planets. The planetary orbits are slightly inclined from the main disk in position angle (PA), but are almost aligned with the inner disk \citep{lagrange_position_2012}. The middle panel shows both orbits as seen from the pole, with the orbit of \Bpic{} b being nearly circular and the orbit of \Bpic{} c slightly eccentric. The separation between both planets and the star is shown in the left panel. In all three panel, the orbit of \Bpic{} b is shown in blue and that of \Bpic{} c is in red, along with 100 random MCMC realisations. The main uncertainty in the trajectory of \Bpic{} b appears to com from a lack of data in the northeastern part of its orbit, leading to a large range of possible orbital periods.}
    \label{fig:architecture}
\end{figure*}

\begin{figure}[ht!]
    \centering
    \includegraphics[width=\linewidth]{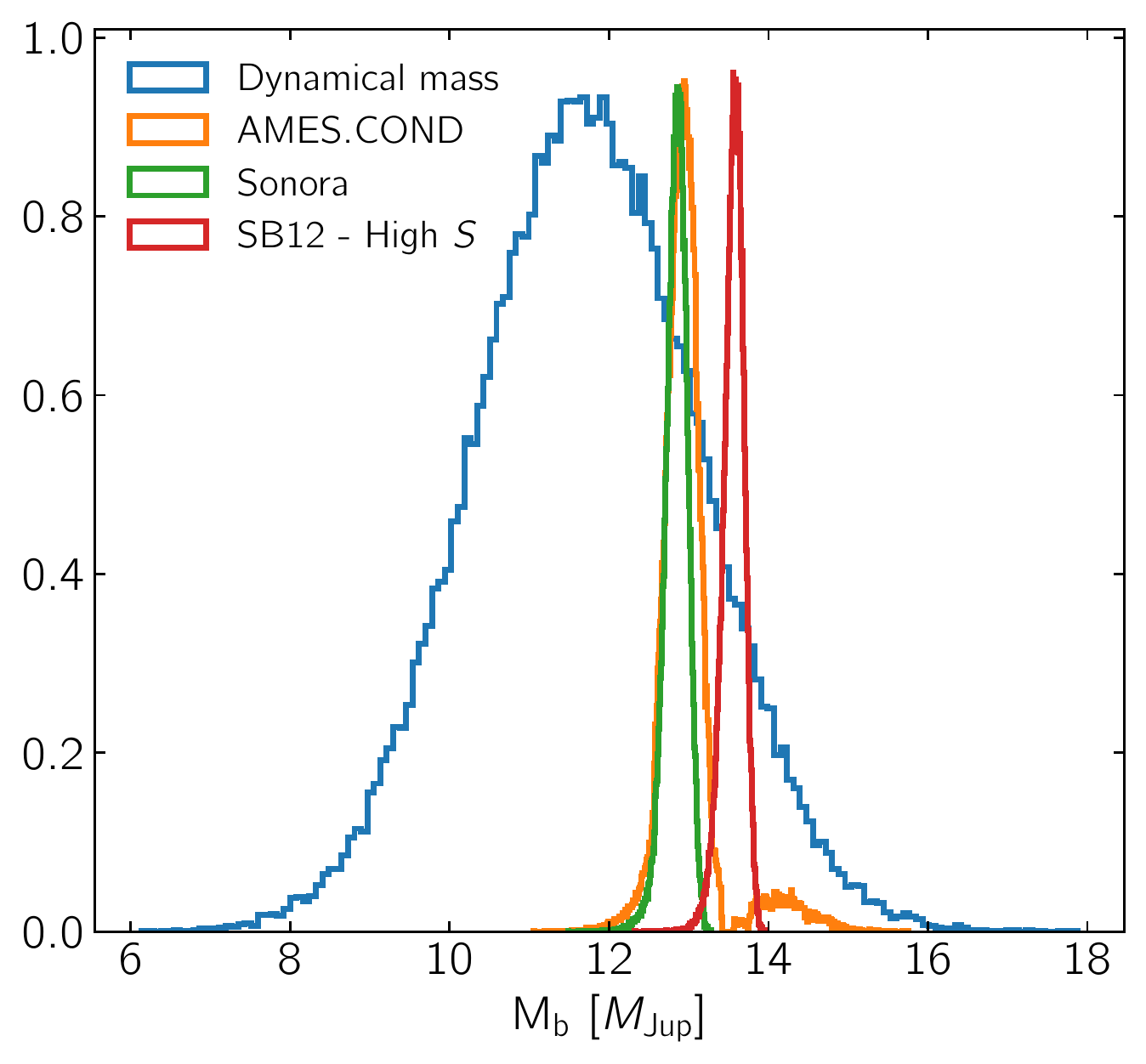}
    \caption{Distributions of the mass of \Bpic~b from our dynamical measurement (black) compared to predictions from the evolutionary models \textsc{COND} \citep{baraffe_evolutionary_2003}, \textsc{Sonora} (Marley et al. in prep) and the warm-start model with $S=13~\mathrm{k_B}/\mathrm{baryon}$ from \citet{spiegel_spectral_2012} assuming an age of $25\pm3~$Myr \citep{messina_rotation-lithium_2016} and a bolometric luminosity of $\log\mathrm{L}/\mathrm{L}_\odot=-3.76\pm0.02$ \citep{chilcote_12.4$upmu$m_2017}.}
    \label{fig:evol_mass}
\end{figure}

We considered the classical hot start evolutionary model grid \textsc{COND} \citep{baraffe_evolutionary_2003}\footnote{The most recent grid from \citet{baraffe_new_2015} does not reach a luminosity that is dim enough at the age of the system.}, the recent hot start grid \textsc{Sonora}\footnote{\href{10.5281/zenodo.1309034}{10.5281/zenodo.1309034}} (Marley et al. in prep.), and the warm-start grid from \citet{spiegel_spectral_2012}. We use grids computed at solar metallicity since the metallicity of \Bpic~b is not well constrained from existing spectrophotometric data, with estimated values from $-0.5$ up to $0.63$ \citep{nowak_peering_2020}, depending on the data and analysis method. The distributions of the predicted mass for all three evolutionary models were derived with a Monte Carlo process from the normal distributions of the most recent luminosity measurement for \Bpic~b of $\log\mathrm{L}/\mathrm{L}_\odot=-3.76\pm0.02$ \citep{chilcote_12.4$upmu$m_2017} and from the age of $25\pm3$Myr \citep{messina_rotation-lithium_2016} (this age assumes \Bpic~b formed very quickly after the birth of the host star). The grid luminosity was first interpolated in $\log t$, at each randomly selected age, for each mass bin. The $\log\mathrm{M}$ was then interpolated in $\log\mathrm{L}_\mathrm{bol}$ at each randomly selected luminosity to derive the model-dependent mass. The process was repeated $10^5$ times to derive the distribution of the model-derived mass. Figure~\ref{fig:evol_mass} shows the distributions for the three models, compared to our dynamical mass measurement. We found $12.9^{+0.2}_{-0.3}~$\Mjup, $12.9\pm0.1~$\Mjup, and $13.6\pm0.1~$\Mjup~ for the \textsc{COND, Sonora} and warm-start model at the highest available entropy, $13~\mathrm{k_B}/\mathrm{baryon}$, following the analysis of \citet{chilcote_12.4$upmu$m_2017}. The results are still consistent with \citet{morzinski_magellan_2015}, \citet{chilcote_12.4$upmu$m_2017} and \citet{nielsen_gemini_2020}, who used different age and/or luminosity estimates ($23\pm3~$Myr, $24\pm3~$Myr, and $26\pm3~$Myr respectively, and $\log\mathrm{L}_\mathrm{bol}/\mathrm{L}_\odot=-3.78\pm0.03$).  Mostly driven by the tight constraint on the bolometric luminosity, the model derived masses exhibit a precision ten times better than our dynamical mass measurement. Yet, the two cold-start model grids are consistent to better than $1\sigma$ to our dynamical mass measurement and the warm-start model from \cite{spiegel_spectral_2012} is consistent within $2\sigma$ at the measured age of the system, favouring formation with the highest initial entropy, following the same conclusion reached by \citet{nielsen_gemini_2020}. Our precision is neither high enough to precisely calibrate the evolutionary tracks, nor to discriminate which of the hot start or warm-start scenarios best fit the parameters of \Bpic{} b, nor to derive the age of the system assuming the models are accurate.

Alternatively, one can assume the evolutionary models predict correct masses for a given luminosity-age pair. Under such hypothesis, one can conversely use our dynamical mass measurement to determine any significant formation delay of \Bpic{} b, with respect to A, according to its measured luminosity \citep{currie_last_2009}. However, the large uncertainty on the dynamical mass only translates into an upper limit on this delay of 12 Myr at $1\sigma$ for all three evolutionary models, which is more than any timescale for giant planet formation. A precision of $0.1$\,\Mjup{} is required to test this scenario.

A continuous and long term monitoring of \Bpic{} b via relative and absolute astrometry, especially with the exquisite precision of VLTI/GRAVITY, and spectroscopy will enable to improve the orbital parameters and the mass precision to revisit the calibration of the evolutionary models and/or the formation delay hypothesis, following the same conclusions as \citet{nielsen_gemini_2020}. 
As seen in Figure~\ref{fig:architecture}, a lot of orbital solutions for \Bpic{} b would be 
excluded by a better constraint on its period. The fact that both relative astrometry and RV data 
cover less than a full period for \Bpic{} leads to a greater uncertainty in $P_\mathrm{b}$. Continuous 
monitoring of \Bpic{} b in the next 5 years will strongly constrain maximum separation of \Bpic{} b 
in the northeastern half of the orbit, hence enabling a better precision in $P_\mathrm{b}$. Furthermore, 
continuous RV monitoring of \Bpic{} as well as the use of unbiased absolute astrometric measurements from \textit{Gaia} in combination with the long baseline with \textit{Hipparcos}  would not only help to constrain the period of \Bpic{} b, but 
it would also improve the accuracy and precision of model-independent mass measurements for both planets. 

Similarly, a direct detection of \Bpic~c together with an adequate sampling of its spectral energy distribution would provide another independent luminosity measurement within the same system, at a lower dynamical mass. This would further strengthen the tests of the evolutionary models. 
Direct imaging of \Bpic{} c may be possible with the \textit{James Webb Space Telescope} (\textit{JWST}) or with 
next-generation giant ground based telescopes such as the European Extremely Large Telescope 
(E-ELT), which should provide an angular resolutions of $\sim 100$ mas for the former 
and $\sim 30$ mas for the latter. As shown in Figure~\ref{fig:architecture}, the separation 
between \Bpic{} c and its host is greater than 100 mas for a significant fraction of its 
orbit according to the solutions derived in this work. Such a measurement would independently 
confirm the existence of the planet while providing another measurement to calibrate 
mass-luminosity models.

\section{Conclusion}\label{sec:conclusion}

We present a new and independent analysis of the RVs of \Bpic{}. For the first time, we use a SHO GP framework to model the stellar activity of the star, with a training of the model from the photometric lightcurve, which results in the following results:
\begin{itemize}
    \item We show that the SHO GP enables to model the $\delta$ Scuti pulsations with amplitudes as high as several hundreds of meters per second;
    \item It yields better precision than a QP kernel typically used to model stellar rotation and is favoured according to the BIC;
    \item It yields RV offsets consistent with parametric models from the literature \citep{lagrange_evidence_2019};
    \item It provides a simpler framework than the parametric models, with three hyperparameters instead of up to 90 parameters;
    \item It allows to model poorly sampled RVs, excluded from previous analyses, hence expanding the baseline by nearly 5 years, as early as 2003.
\end{itemize}

This extended RV coverage and the precision of the SHO GP offsets make it possible to constrain 
the mass of both planets using RV data only with low sensitivity to the choice of priors imposed 
on the orbit of \Bpic{} b. However, the constraints on other orbital parameters are sensitive to 
changes in the priors from oribtal fits using relative and absolute astrometry in the literature
\citep{lagrange_post-conjunction_2019,nielsen_gemini_2020}. To better characterize the entire 
system, we also perform a fit of all available relative astrometry and RV data. The joint fits 
and the fits with RV data only yield results that are generally consistent. We find a
model-independent mass of $12.7\pm1.3$~\Mjup{} for \Bpic{} b. A low-eccentricity orbit 
($e=0.016^{+0.044}_{-0.012}$) with a relatively short period of $P_\mathrm{b}=20.6^{+1.5}_{-0.4}$~yr 
is preferred. For \Bpic{} c, the mass is constrained to $8.5\pm0.5$~\Mjup{} and the orbit is 
consistent with values from the literature \citep{lagrange_evidence_2019,nielsen_gemini_2020}.

The dynamical mass measurement of \Bpic{} b presented in this work is consistent with hot-start 
and warm-start evolutionary models at the highest computed initial entropy ($S=13~\mathrm{k_B}/\mathrm{baryon}$) but is not precise enough to discriminate against any of the 
models tested, reaching conclusions similar to \cite{nielsen_gemini_2020}. Further monitoring with 
relative and absolute astrometry, as well as RV, will be required to obtain stronger 
constraints on the mass, enabling a better calibration of mass-luminosity models.

\Bpic{} c has not yet been observed directly, but according to the orbital solutions presented 
here, it could be accessible by direct imaging in the near future with \textit{JWST}. With a minimum mass of $8.5\pm0.5$~\Mjup{}, the COND evolutionary model predicts a contrast with the star around $10.7$ magnitudes and $8.9$ magnitudes, at K-band and L-band respectively. Such a detection would provide 
a second point for calibration of evolutionary model while further improving the 
characterization of orbits in the \Bpic{} system.

This work demonstrates that precise modelling of complex stellar activity with SHO GP is effective. It opens a new path towards the detection of giant planets with RV only around young stars as well as the characterization of systems with directly imaged planets beyond that of \Bpic{}.

\acknowledgments
We sincerely thank the referee for the thoughtful and careful review, and the suggestions, leading to an improvement of the present work. We thank Anne-Marie Lagrange for providing all published individual RV measurements and Djamel M\'ekarnia for the ASTEP light curve. We thank Lauren Weiss for her contribution to early phases of this project and Aur\'elien Wyttenbach and Raphaëlle Haywood for fruitful discussions about GP. TV acknowledges support from an Institut de Recherche sur les Exoplan\`etes (iREx) Trottier Excellence Grant. JR and RD acknowledge support from Fonds de Recherche du Qu\'ebec since JR's work was performed in part under contract with the University of Montr\'eal. 
JR is supported by the French National Research Agency in the
framework of the Investissements d’Avenir program (ANR-15-IDEX-02),
through the funding of the "Origin of Life" project of the Univ.
Grenoble-Alpes."

%% Facilitiies
%% E.g.: HST(STIS)
\vspace{5mm}
%\facilities{Archival data}

%% Software
%% E.g.: Cloudy \citep{2013RMxAA..49..137F}, 
\software{emcee \citep{foreman-mackey_emcee:_2013}, 
celerite \citep{foreman-mackey_fast_2017}, 
RadVel \citep{fulton_radvel:_2018}, 
Astropy \citep{the_astropy_collaboration_astropy_2013}, 
corner \citep{foreman-mackey_corner.py:_2016}, 
orbitize! \citep{blunt_orbitize_2020}.}

\clearpage
\appendix

\section{Posterior Distributions for all Fits}\label{app:all_post}

\begin{figure*}[ht!]
    \centering
    \includegraphics[width=0.95\linewidth]{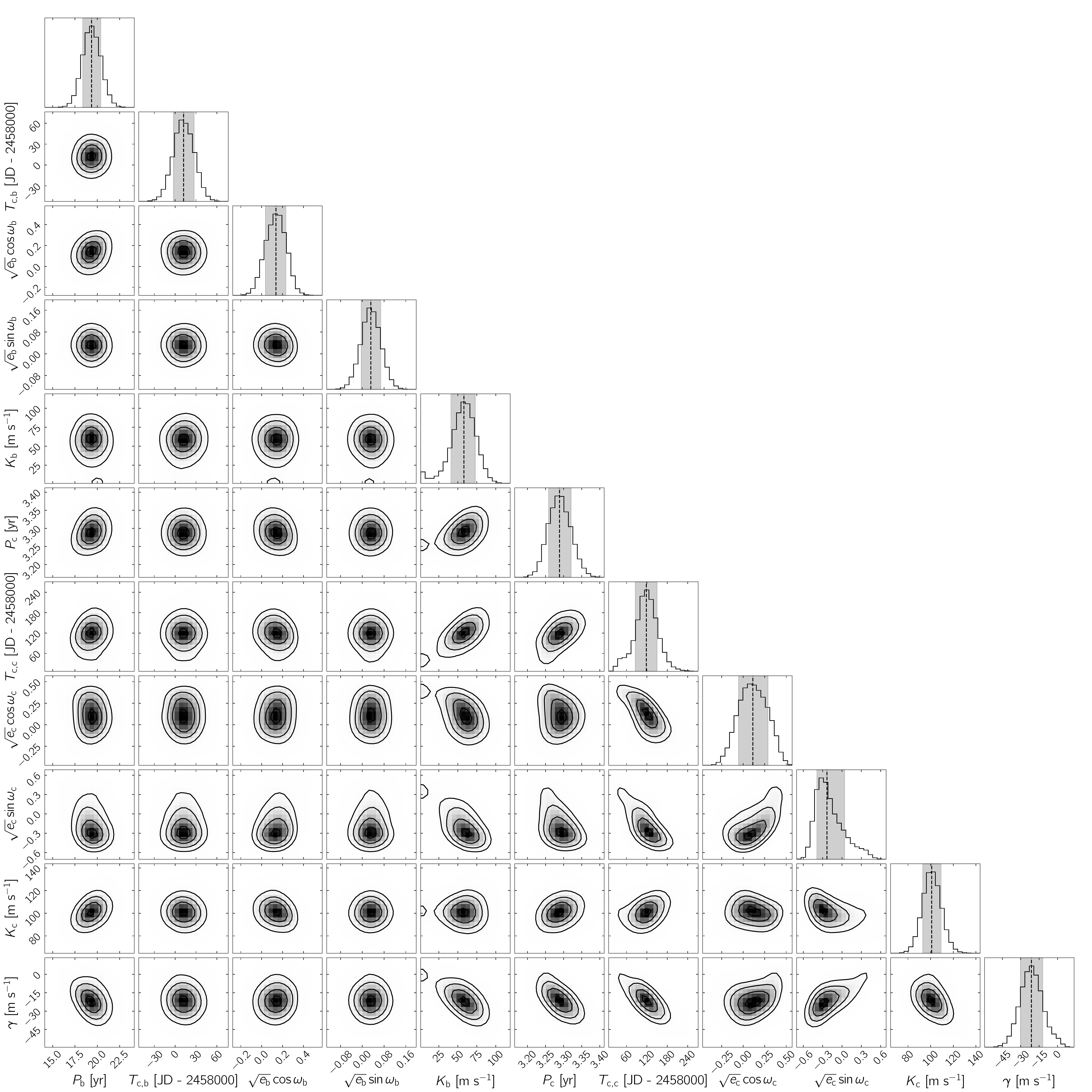}
    \caption{Posterior distribution of the Case A RV fit. The dashed lines and grey shaded areas show the median and the $1\sigma$ interval for each parameter. A small bimodality in $M_\mathrm{b}$ suggsets that the orbit of \Bpic{} b is not totally well-constrained by the 2008-2018 RV offsets only.}
    \label{fig:corner_a}
\end{figure*}

\begin{figure*}[ht!]
    \centering
    \includegraphics[width=0.95\linewidth]{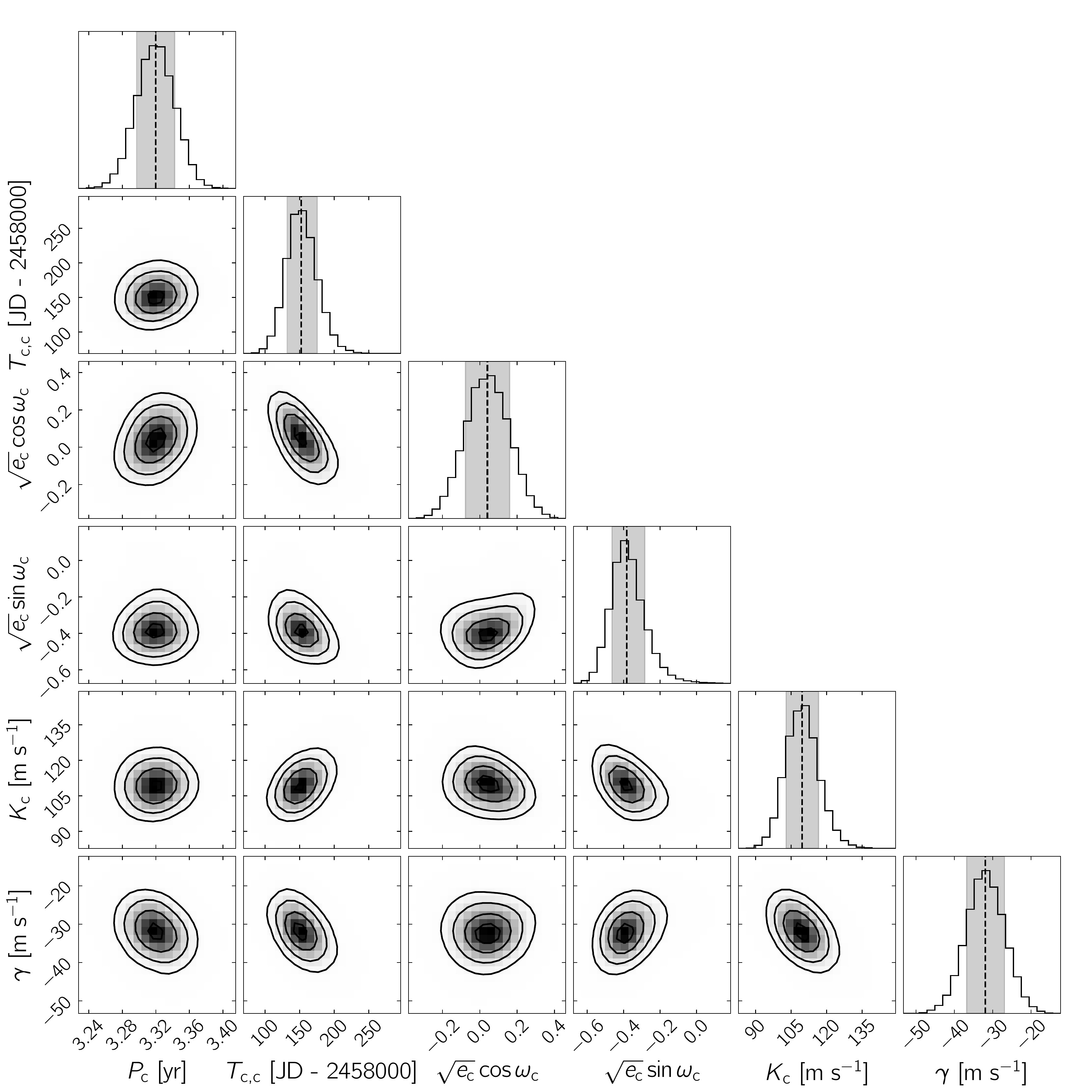}
    \caption{Posterior distribution of the Case B RV fit. The dashed lines and grey shaded areas show the median and the $1\sigma$ interval for each parameter. The orbit of \Bpic{} b is fixed to values from \cite{lagrange_post-conjunction_2019}.}
    \label{fig:corner_b}
\end{figure*}

\begin{figure*}[ht!]
    \centering
    \includegraphics[width=0.95\linewidth]{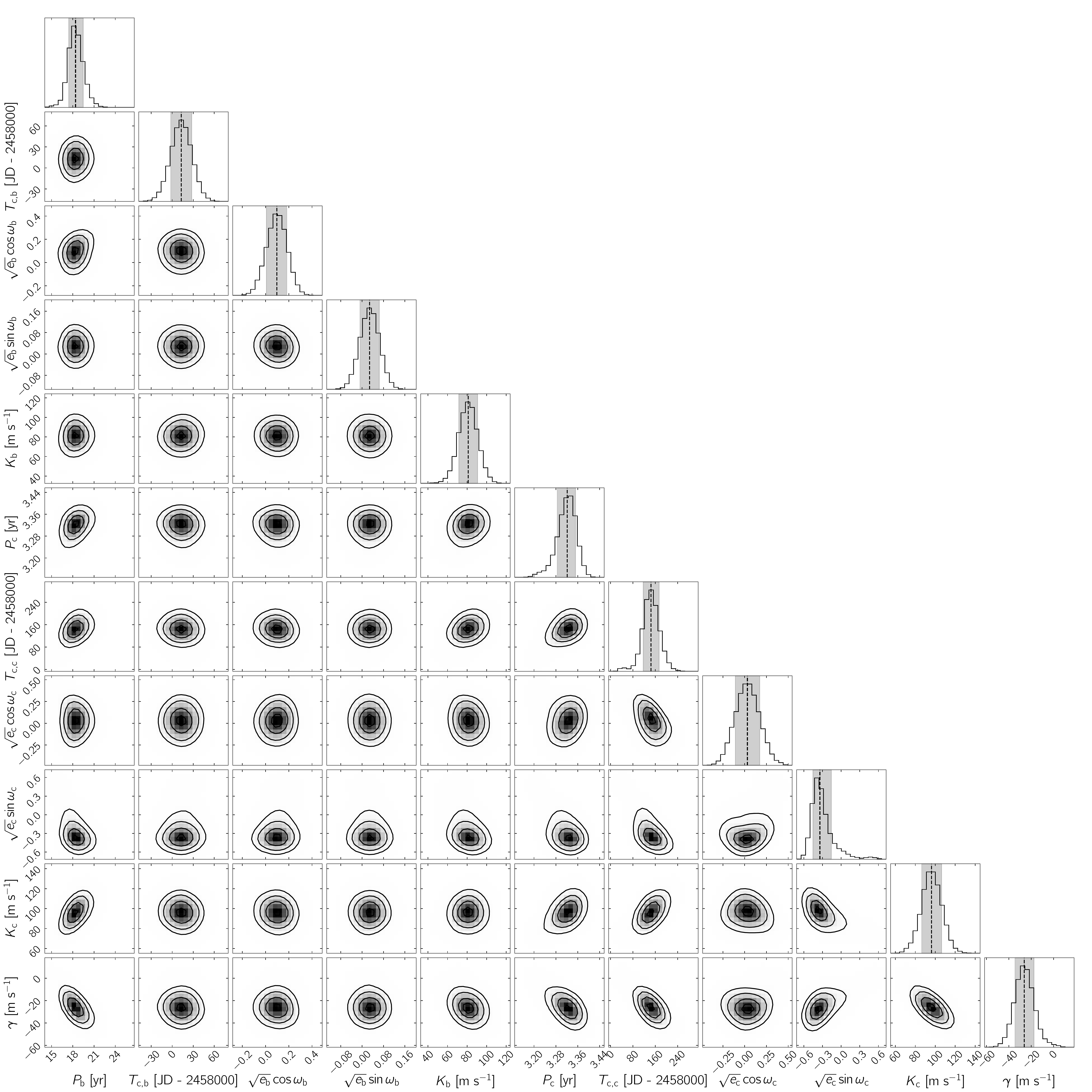}
    \caption{Posterior distribution of the Case C RV fit. The dashed lines and grey shaded areas show the median and the $1\sigma$ interval for each parameter. Gaussian priors from \citet{lagrange_post-conjunction_2019} were applied on the orbital parameter of \Bpic{} b, but with a wide prior of $22 \pm 4$ yr on $P_\mathrm{b}$.}
    \label{fig:corner_c}
\end{figure*}

\begin{figure*}[ht!]
    \centering
    \includegraphics[width=0.95\linewidth]{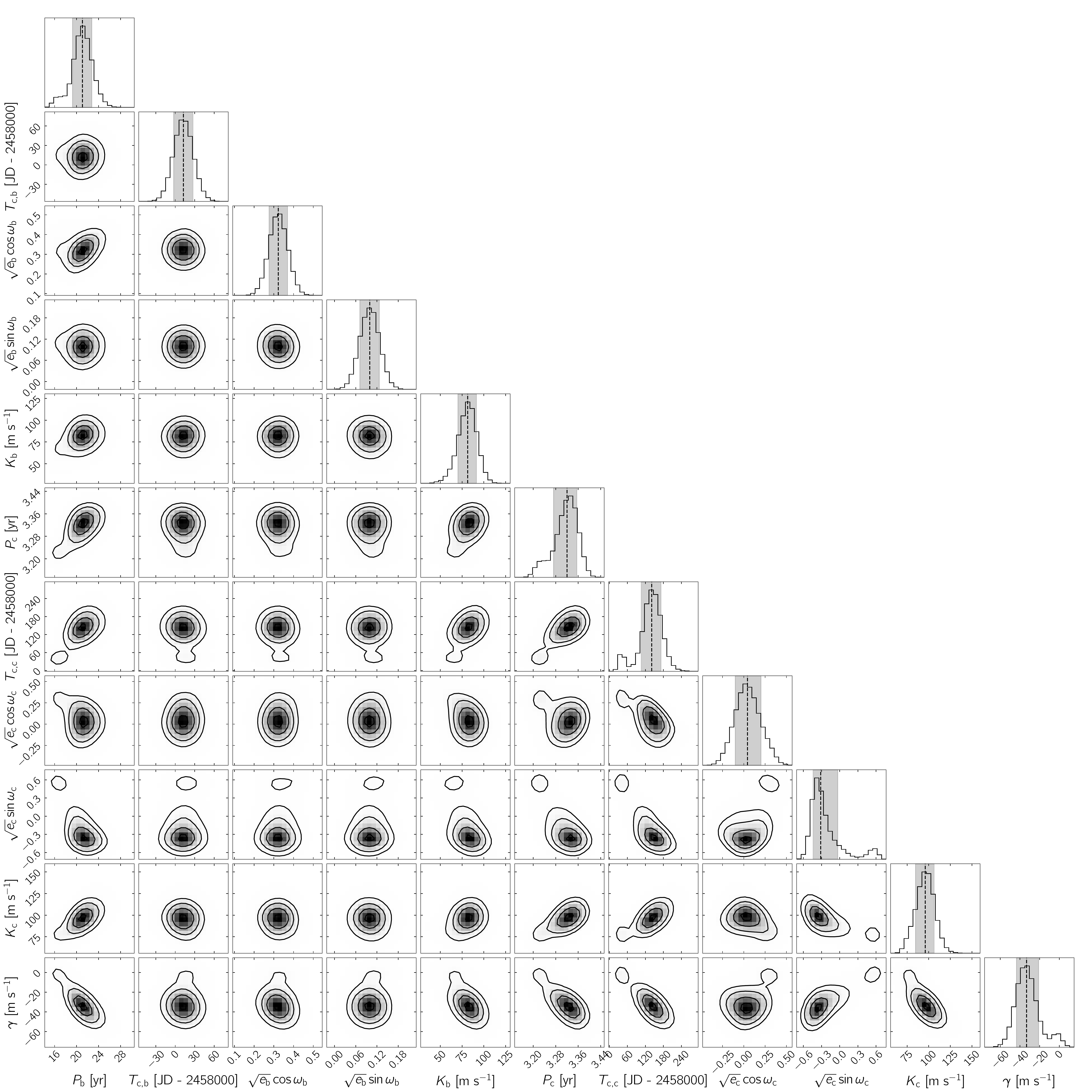}
    \caption{Posterior distribution of the Case D RV fit. The dashed lines and grey shaded areas show the median and the $1\sigma$ interval for each parameter. Gaussian priors from \citet{nielsen_gemini_2020} were applied on the orbital parameter of \Bpic{} b, but with a wide prior of $22 \pm 4$ yr on $P_\mathrm{b}$.}
    \label{fig:corner_d}
\end{figure*}

\begin{figure*}[ht!]
    \centering
    \includegraphics[width=0.95\linewidth]{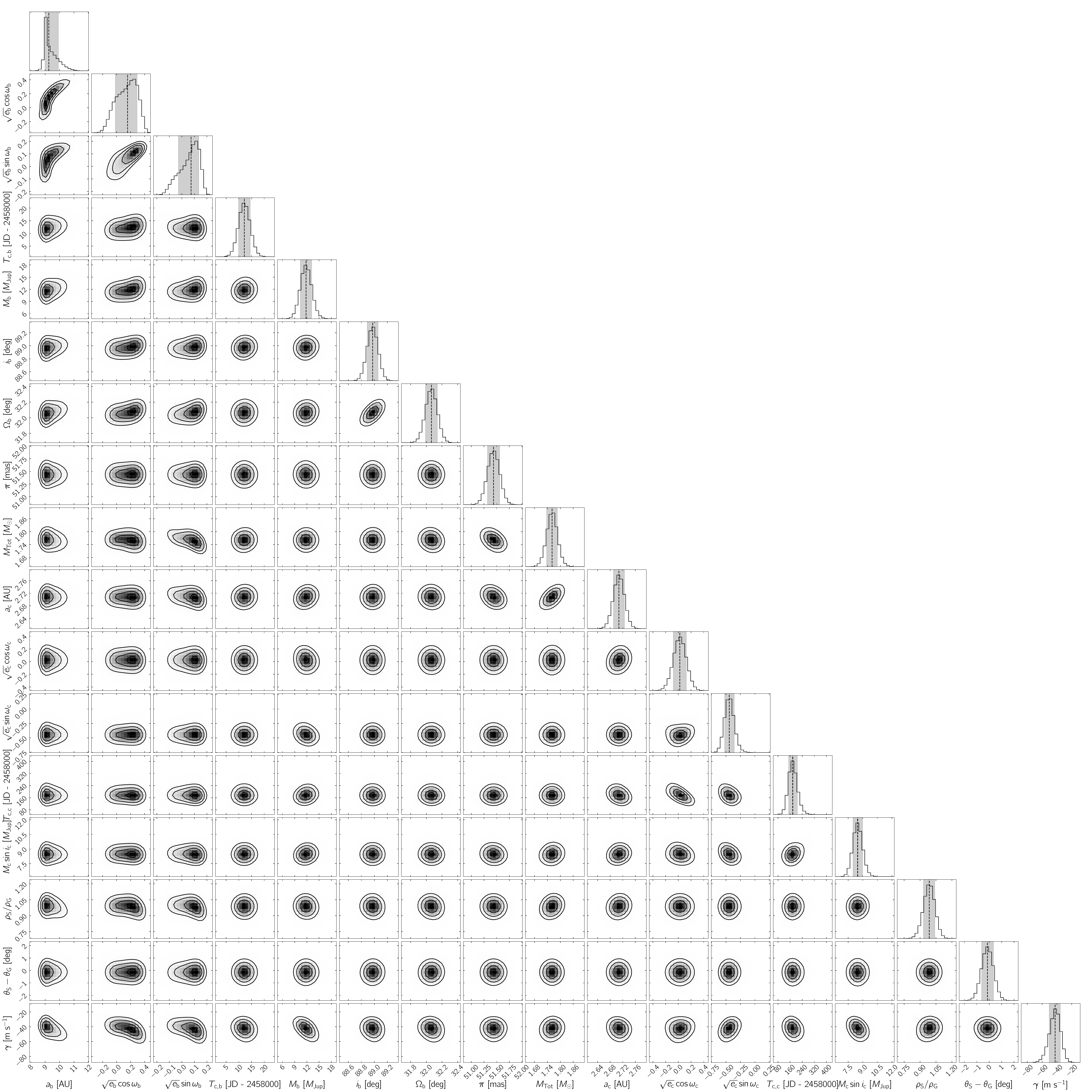}
    \caption{Posterior distribution for the joint fit of RV and relative astrometry, without the VLTI/GRAVITY measurement of September 2008. The dashed lines and grey shaded areas show the median and the $1\sigma$ interval for each parameter. We note very strong correlations between $a_\mathrm{b}$, $h_\mathrm{b} = \sqrt{e_\mathrm{b}}\cos{\omega_\mathrm{b}}$, $k_\mathrm{b} = \sqrt{e_\mathrm{b}}\sin{\omega_\mathrm{b}}$.}
    \label{fig:corner_no_gravity}
\end{figure*}

\begin{figure*}[ht!]
    \centering
    \includegraphics[width=0.95\linewidth]{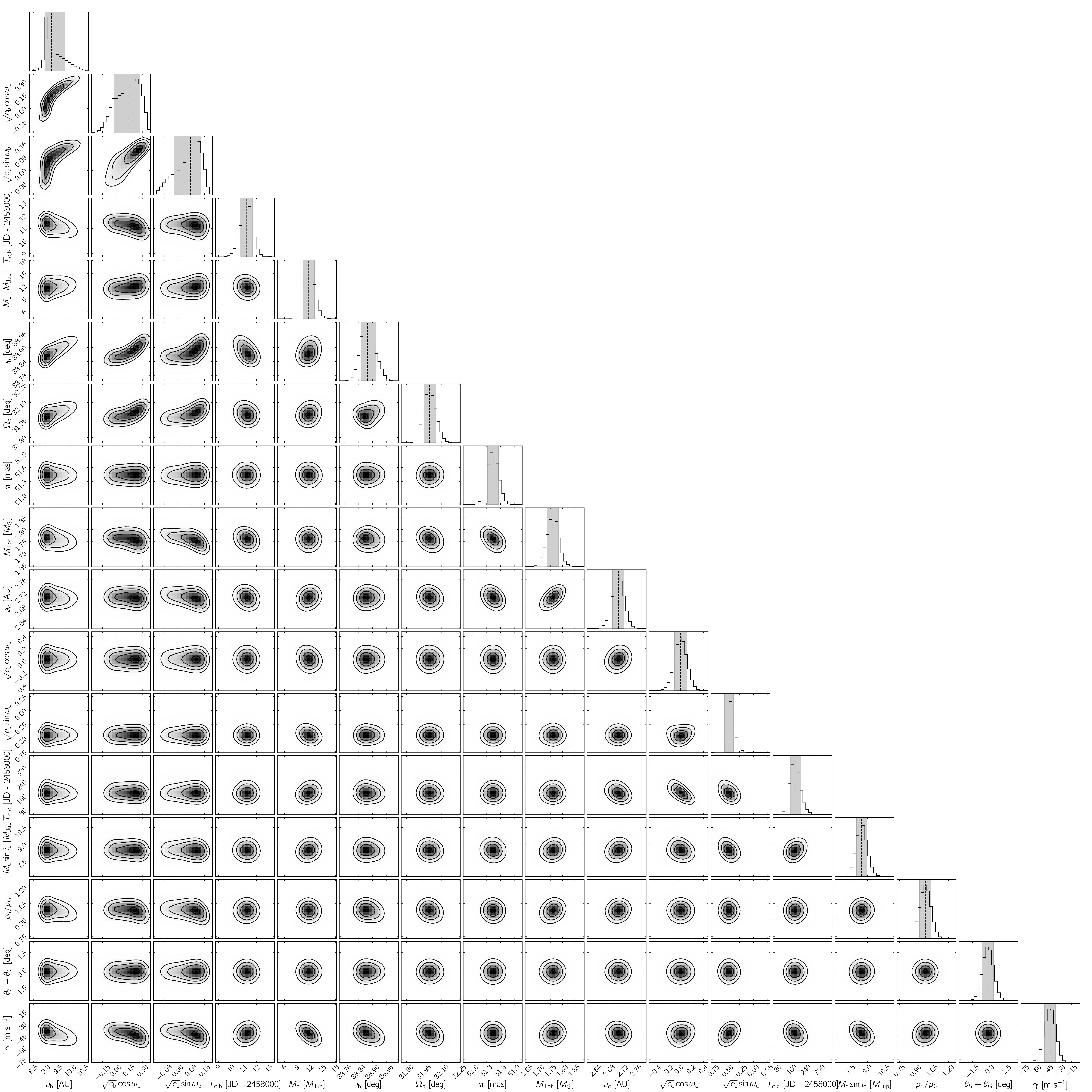}
    \caption{Posterior distribution for the joint fit of RV and relative astrometry, with the VLTI/GRAVITY measurement of September 2008. The dashed lines and grey shaded areas show the median and the $1\sigma$ interval for each parameter. As in Figure~\ref{fig:corner_no_gravity}, $a_\mathrm{b}$, $h_\mathrm{b} = \sqrt{e_\mathrm{b}}\cos{\omega_\mathrm{b}}$, $k_\mathrm{b} = \sqrt{e_\mathrm{b}}\sin{\omega_\mathrm{b}}$ are strongly correlated.}
    \label{fig:corner_gravity}
\end{figure*}

\clearpage

\section{Table with newly derived RV offset + error}\label{app:RV_data}

\begin{table*}[ht!]
    \caption{RV Offsets from 2003-2018.}\label{tab:rv_offsets}
    \begin{center}
    \begin{tabular}{cccc} 
        \hline\hline             
        MJD & RV & $\sigma_{\mathrm{RV}}$ \\
            & [\mps{}] & [\mps{}]\\
        \hline  
        \hline
        52944.376672 & -77.43  & 59.87  \\
        52993.188574 & 39.68   & 28.94  \\
        53048.602259 & -154.64 & 34.06  \\
        53775.641184 & -9.07   & 149.54 \\
        54056.325773 & 20.93   & 106.71 \\
        54112.579919 & -99.04  & 247.39 \\
        54441.746852 & -191.08 & 148.94 \\
        54483.110698 & -92.49  & 135.89 \\
        54544.046717 & -76.74  & 16.72  \\
        54780.215069 & -70.24  & 24.16  \\
        54799.213508 & -71.54  & 25.30  \\
        54829.165935 & -202.14 & 27.49  \\
        54850.142609 & -130.78 & 22.24  \\
        54913.040806 & -84.29  & 18.44  \\
        55131.298795 & 2.44    & 26.38  \\
        55170.202032 & 18.54   & 22.49  \\
        55519.296681 & 37.59   & 30.68  \\
        55566.194698 & 148.72  & 14.54  \\
        55597.126002 & 85.80   & 19.29  \\
        56569.311461 & 129.29  & 41.54  \\
        56582.333626 & 95.49   & 28.91  \\
        56685.203668 & 24.62   & 32.80  \\
        56694.067479 & 91.60   & 27.57  \\
        56706.582371 & 198.96  & 23.64  \\
        56984.282332 & -15.44  & 20.36  \\
        57344.247022 & -148.18 & 55.92  \\
        57667.303333 & 106.83  & 22.29  \\
        57712.283914 & 128.33  & 21.42  \\
        57771.166012 & 81.25   & 8.54   \\
        57849.014450 & 151.55  & 25.00  \\
        58007.347048 & 3.42    & 21.89  \\
        58037.271945 & 37.74   & 7.34   \\
        58064.265248 & -62.63  & 24.20  \\
        58094.192294 & -35.32  & 6.94   \\
        58207.056451 & -32.82  & 11.04  \\
        58235.001291 & -70.57  & 7.73   \\
        \hline
        \end{tabular}
        \end{center}
\end{table*}

\clearpage

%% Bibliography with aasjournal sytle
\bibliography{bpic.bib}
\bibliographystyle{aasjournal}

% \listofchanges
\end{document}